\documentclass[a4paper,11pt]{article}
\pdfoutput=1 

\usepackage{jcappub} 

\usepackage[T1]{fontenc} 
\usepackage{amsmath}
\usepackage{amssymb}
\usepackage{gensymb}
\usepackage[referable]{threeparttablex}
\usepackage{booktabs, longtable}

\usepackage{grffile}  

\newcommand{\be}{\begin{equation}}
\newcommand{\ee}{\end{equation}}
\newcommand{\diff}{\ensuremath{\mathrm{d}}}
\newcommand{\Rin}{\ensuremath{R^\prime}}
\newcommand{\Zin}{\ensuremath{Z^\prime}}

\usepackage{xcolor}

\definecolor{darkBlue}{rgb}{0, 0, 0.8}

\definecolor{dark-red}{rgb}{0.0,0.0,0.0}
\definecolor{green}{rgb}{0.0,0.0,0.0}
\definecolor{DarkBlue}{rgb}{0.0,0.0,0.0}

\def\Xmax{$X_{\max}$ }
\def\X2{Var($X_{\max})$}

\renewcommand{\epsilon}{\varepsilon}

\title{Impact of the finite life-time of UHECR sources}

\author[a,b,c,1]{B.~Eichmann\note{Corresponding author.}}
\author[a]{and M.~Kachelrie\ss}

\affiliation[a]{Norwegian University for Science and Technology (NTNU), Institutt for fysikk, Trondheim, Norway}

\affiliation[b]{Ruhr-Universit\"at Bochum, Theoretische Physik IV, Fakult\"at f\"ur Physik und Astronomie, Bochum, Germany}

\affiliation[c]{Ruhr Astroparticle and Plasma Physics Center (RAPP Center), Bochum, Germany}

\keywords{ultrahigh energy cosmic rays, radio galaxies}

\abstract{
The observational data on ultrahigh energy cosmic rays (UHECR), in particular
their mass composition, show strong indications for extremely hard spectra of
individual mass groups of CR nuclei at Earth. In this work, we show that
such hard spectra can be the result of the finite life-time of UHECR sources,
if a few individual sources dominate the UHECR flux at the highest energies.
In this case, time delays induced by deflections in the turbulent extragalactic
magnetic field as well as from the diffusive or advective escape from the
source environment can suppress low-energy CRs, leading to a steepening
of the observed spectrum.
Considering radio galaxies as the main source of UHECRs, we discuss the
necessary conditions that few individual sources dominate over the total
contribution from the bulk of sources that have been active in the past.
We provide two proof-of-principle scenarios showing that for a turbulent
extragalactic magnetic field with a strength $B$ and a coherence length
$l_{\rm coh}$, the  life-time of a  source at a distance $d_{\rm src}$
should satisfy
${t_{\rm act} \sim \left( B/1\,\text{nG} \right)^2\,\left( d_{\rm src}/10\,\text{Mpc} \right)^2\,\left( l_{\rm coh}/1\,\text{Mpc} \right)\,\text{Myr}}$ to
obtain the necessary hardening of the CR spectrum at Earth.
}

\begin{document}
\maketitle
\flushbottom

\section{Introduction}
\label{sec:intro}

The quest to reveal the sources of the highest energy particles observed has
been a key-driver in the field of cosmic ray (CR) physics since the first
observation of CRs with energy close to $10^{20}$\,eV~\cite{Linsley:1963km}.
On the experimental side, considerable progress has been made in the
last 15~years as a result of the observations made with the Pierre
Auger Observatory (PAO) and the Telescope Array (TA) experiment, for
recent reviews see~\cite{Anchordoqui:2018qom,Kachelriess:2019oqu,Kachelriess:2022phl}. In
particular, the diffuse energy spectrum~\cite{TelescopeArray:2012qqu,PierreAuger:2020qqz,PierreAuger:2020kuy} of ultrahigh energy cosmic rays (UHECRs)
is now known rather precisely, and several deviations from a simple power law
have been determined: In addition to the second knee around
$E\simeq 5\times10^{17}$\,eV,
the ankle at $E\simeq5\times 10^{18}$\,eV and the flux suppression at the
highest energies, a fourth feature, the so-called instep, has been
established by the PAO~\cite{PierreAuger:2020kuy,PierreAuger:2020qqz}.
The second important piece of information, the mass composition of the
UHECR flux, can be inferred only indirectly from the number of particles
an air shower is containing as function of atmospheric depth $X$,
relying on simulations for hadronic interactions. 
Comparing the predicted \Xmax distributions for a mixture of CR nuclei to
the observed  distribution, one can fit the relative fraction of the  CR
nuclei: Above $10^{18}$\,eV, the dominant component in the UHECR flux changes
successively from protons, to helium and nitrogen, a behaviour suggestive
for the presence of a Peters' cycle. According to these fits, the different
dominating element groups are surprisingly well
separated~\cite{PierreAuger:2020kuy}.

Adding  some assumptions on the propagation of UHECRs (e.g., a
model for the extragalactic magnetic field) allows one to
test if specific UHECR source models can reproduce these data.
The arguable simplest theoretical model employs a continuous, uniform
distribution of identical sources, combined often with a negligible
extragalactic magnetic field (EGMF).
Inspired by the idea of a Peters' cycle, a rigidity dependent source
spectrum with an exponential cut-off,
$\diff N_i/\diff R = K_i  R^{-\alpha} \exp(-R/\hat{R})$, is typically
used~\cite{Taylor:2015rla,Auger2017,Romero-Wolf:2017xqe,AlvesBatista:2018zui,Heinze:2019jou,Guido:20218S}. Fitting then the common slope $\alpha$ and
cut-off $\hat{R}$ as well as the relative abundances $K_i$ of the
different mass groups, a satisfactory description of both the spectrum
and the composition data can be achieved despite the simplicity of this
model. However, in order to obtain a clean separation of the mass groups
and thus to reproduce the composition data,
a very hard slope of the injection
spectrum into extragalactic space is required: For instance, the fits 
presented in Ref.~\cite{Guido:20218S} use slopes\footnote{Note the sign.}
between $\simeq -1$ and $-2$. Moreover, the
ankle has to be explained by a second extragalactic population which does
not show the same Peters' cycle but has a dominantly light composition.
Alternatively, there are also some approaches~\cite{Taylor:2015rla,Auger2017} that used a slope much closer to 2 as well as a dominant contribution from (identical) local sources. But in these approaches, only the data
  above the ankle are taken into account yielding the need of a sharp flux suppression of the sub-ankle contribution to keep the variance of the chemical composition low at about 10\,EeV. Still, these results show some discrepancies with the data, which however, is unavoidably as there is necessarily a mixture of different CR nuclei types at about 10\,EeV in the case of an initial $1/E^2$ spectrum.

Since the assumption of identical sources is not well justified, one may
ask how these fits change if one uses a source populations with, e.g.,
a distribution of $\hat{R}$ values. Considering only the energy spectrum,
it is clear that a distribution of maximal rigidities $\hat{R}$ 
decreasing faster than  $1/\hat{R}$ will require even harder spectra of
individual sources~\cite{Kachelriess:2005xh,Kachelriess:2022phl}. Moreover, a distribution
of $\hat{R}$ values will spread out the different mass groups,
even if they are well separated in the spectra of individual sources.
This qualitative argument was quantified in the recent work~\cite{Ehlert+2022},
which showed that UHECR sources have to be close to ``standard candles'',
if many sources contribute to the flux above $10^{18}$\,eV.

One may also wonder how such hard escape spectra may be generated.
While first-order Fermi acceleration \cite{1977DoSSR.234.1306K, Bell1978a, Bell1978b, 1977ICRC...11..132A, 1978ApJ...221L..29B} predicts for  strong, non-relativistic
shocks in the test-particle approximation  the slope $\alpha\simeq 2$ for the
source spectrum,  the slope is modified by  non-linear and relativistic effects.
However, the expected range of slopes arising from these modifications 
is far from the values $-2\lesssim\alpha\lesssim -1$ obtained in the fits.
Alternative acceleration mechanisms, as e.g.\ unipolar acceleration
  in pulsars, may however lead naturally to flatter energy
  spectra~\cite{Fang:2012rx}.

Another possibility is that an energy-dependent escape from the shock
modifies strongly the spectra. In the case of non-relativistic shocks
around supernova remnants,
it was however argued that an acceleration spectrum that is flatter than
$1/E^2$ will result in an $1/E^2$ escape spectrum---thus this
effect works in this case in the opposite direction~\cite{Schure:2013yga}.
A final option is the modification of spectra by threshold effects of
photo-nuclear $A\gamma$ interactions in the source, see, e.g.,
Refs.~\cite{Globus:2014fka,Unger:2015laa,Kachelriess:2017tvs,Biehl:2017zlw}.

In this work, we suggest to use the magnetic horizon
  effect~\cite{Parizot:2004wh,Lemoine:2004uw,Berezinsky:2005fa} applied to a
  small number of source with finite life-time as explanation
for the apparently hard injection spectra obtained in fits to the observed
spectrum and composition of UHECR data.
In its original version,  the magnetic horizon  effect was used in
Refs.~\cite{Parizot:2004wh,Lemoine:2004uw,Berezinsky:2005fa} to explain
the suppression of the diffuse extragalactic cosmic ray flux below
few\,$\times 10^{17}$\,eV in the ``dip model''~\cite{Berezinsky:2002nc}.
Later, the same effect was studied for the case of a mixed composition:
For instance, Ref.~\cite{Mollerach:2013dza} showed that the suppression of
low-energy cosmic rays helps to reconcile the measured composition data
with steeper injection spectra.
More recently, Ref.~\cite{Mollerach:2019wne} studied the same effect for
a single source in the local Supercluster.
Our starting point in this work is the observation
made in Ref.~\cite{Eichmann:2022ias} that a very small number of
local radio galaxies can dominate the UHECR flux above the ankle.
In this scenario, the instep is just the most
obvious of several irregularities which are caused by the small number of
sources contributing to the high-energy end of the energy spectrum.
Since in this model, UHECRs are accelerated in the extended jets
of radio galaxies, a modification of the escape
spectra by $A\gamma$ interactions cannot be used to explain
the hard escape spectra. However, the small number of sources
opens up the possibility that the finite life-time of these
sources together with energy-dependent time delays in the
extragalactic magnetic field modifies the observed spectra.
In particular, we show that the delay of low-energy CRs leads to a
flattening of the energy spectra of these sources, which can reconcile
the composition data with a standard acceleration spectrum close to
$1/E^2$. 

This paper is structured as follows: In Sec.~\ref{sec:model}, we first
introduce the model used by us for the CR escape and the time evolution
of the CR luminosity of individual sources. In addition, we present how
we treat the CR contribution from the large-scale distribution of radio
sources that has been active in the past. In Sec.~\ref{sec:ind-to-csf}, we
investigate the constraints on the currently active, local sources in order
that they dominate the UHECR flux above the ankle. In Sec.~\ref{sec:compConstr},
we provide some proof-of-principle examples, where we account for a single
local source with a finite life-time to explain the observed energy spectrum
and the composition data. Finally, we discuss in Sec.~\ref{sec:disc&concl}
our results and provide some conclusions.

\section{Cosmic ray sources with a finite life-time}
\label{sec:model}

Active galactic nuclei (AGN) show a varied and complex
history~\cite{Konar+2019,Maccagni+2020,Croston:2009zc},
so that the actual source properties can significantly deviate from the
time-integrated ones. For variations on sufficiently short time scales,
one may expect that time-delays due to diffusion in magnetic fields
smooth out the source variability in the UHECR flux. Here, we will
concentrate on the variability on the largest time scales, which we
associate with the life-time $t_{\rm act}$ of the source, which is
of order of 1--100\,Myr in the case of radio galaxies. 

In terms of the other characteristics of the CR ejecta of radio galaxies, we
stick to the approach that has been introduced in previous
works~\cite{Eichmann+2018, Eichmann2019, Eichmann:2022ias} and suppose that the
fraction $g_{\rm m}<1$ of the jet power $L_{\rm jet}$ is transferred to CRs,
while the radio-to-jet power correlation determines
$L_{\rm jet}\propto L_{151}^{\beta_{\rm L}}$, where in general
$0.5\lesssim \beta_{\rm L}\lesssim 0.9$ is observed~\cite{Ineson+2017, GodfreyShabala2016, GodfreyShabala2013, Willott+1999}. Unless otherwise stated,
we will subsequently adopt a rather strong dependence with $\beta=0.89$ as
suggested from one of the most recent works~\cite{Ineson+2017}.
Further, the associated maximal rigidity of CRs is then given by
$\hat{R}=g_{\rm acc} \sqrt{(1-g_{\rm m})L_{\rm jet}/c}$, where the acceleration
efficiency $0.01\lesssim g_{\rm acc}\lesssim 1$ encapsulates all details of
the acceleration process.

\subsection{Time-dependent luminosity}

We will model the time variability of a UHECR source which
in principle can affect the energy spectrum in a complicated way
by a sole parameter, the overall CR luminosity  $L_{\rm cr}$. In the
simplest picture, implementing a finite life-time as a step function,
$L_{\rm cr}$ vanishes if the activity time satisfies
$t_{\rm act}<|t_{\rm obs}-[t(\epsilon)-d_{\rm src}/c]|$
for a source at distance $d_{\rm src}$ and the travel time $t(\epsilon)$
of a CR with energy $\epsilon$. More realistically, the CR luminosity will
evolve during the life-time of the source. To be specific, we choose this
time dependence as a Gaussian normal distribution,
\be
L_{\rm cr}(t^\prime) =
\frac{L_0}{\exp\left[ - (t_{\rm max}/\sigma_{\rm act})^2/2 \right]}\,
\exp\left[ - \frac{1}{2}\, \left(\frac{t^\prime+d_{\rm src}/c-t_{\rm max}}{\sigma_{\rm act}}\right)^2\right]\,,
\ee
where $L_0$ denotes the present CR luminosity of the source measured by
the observer, i.e.\ at the time $t^\prime=-d_{\rm src}/c$ with respect to the
observation time $t_{\rm obs}$. Further, $t_{\rm max}$ is the time of maximal
luminosity and the source's active time is connected to the  variance as
$\sigma_{\rm act}=t_{\rm act}/3$. Thus, we define the activity time of a source
as the time in which 99,7\% of the total ejected CR energy is emitted. 
Note that for $-t_{\rm act} \lesssim t_{\rm max}\ll 0$, the source showed a much
higher CR luminosity in the past that increases for a decreasing difference
of $t_{\rm act}-|t_{\rm max}|$. 
Moreover, there is an additional modification of the actually observed
source luminosity,
if the CRs escape the source environment by diffusive or advective transport
as illustrated in the following.

\subsection{Escape from the source environment}

We assume that UHECRs are accelerated by shocks in the jets or lobes of
the AGN, at large distances $\gtrsim 1\,\text{kpc}$ from its central engine. 
At a distances of about $10\,\text{kpc}$ from the central nucleus the photo-hadronic energy loss due to the presence of a strong AGN photon field (with a bolometric luminosity of $\sim 10^{46}\,\text{erg s}^ {-1}$) occurs already on a timescale $\gtrsim 100\,\text{Myr}$ for CR energies $\lesssim 10^{19}\,\text{eV}$. In addition, proton synchrotron losses occur on a timescale \cite{Aharonian2002} $\tau_{\rm syn,p}\sim 14\,(E/10^{19}\text{eV})^{-1}\,(B_{\rm jet}/1\text{mG})^{-2}\,\text{Myr}$, so that we obtain similar timescales for a realistic magnetic field strength of $B_{\rm jet}\sim 0.1\,\text{mG}$ at these energies. Hence, at these distances from the central nucleus energy losses can be neglected for escape times $\lesssim 100\,\text{Myr}$, since even at the highest energies CRs escape the source environment before loosing a significant fraction of their energy.

\begin{figure}[htbp]
\centering
\includegraphics[width=.43\linewidth]{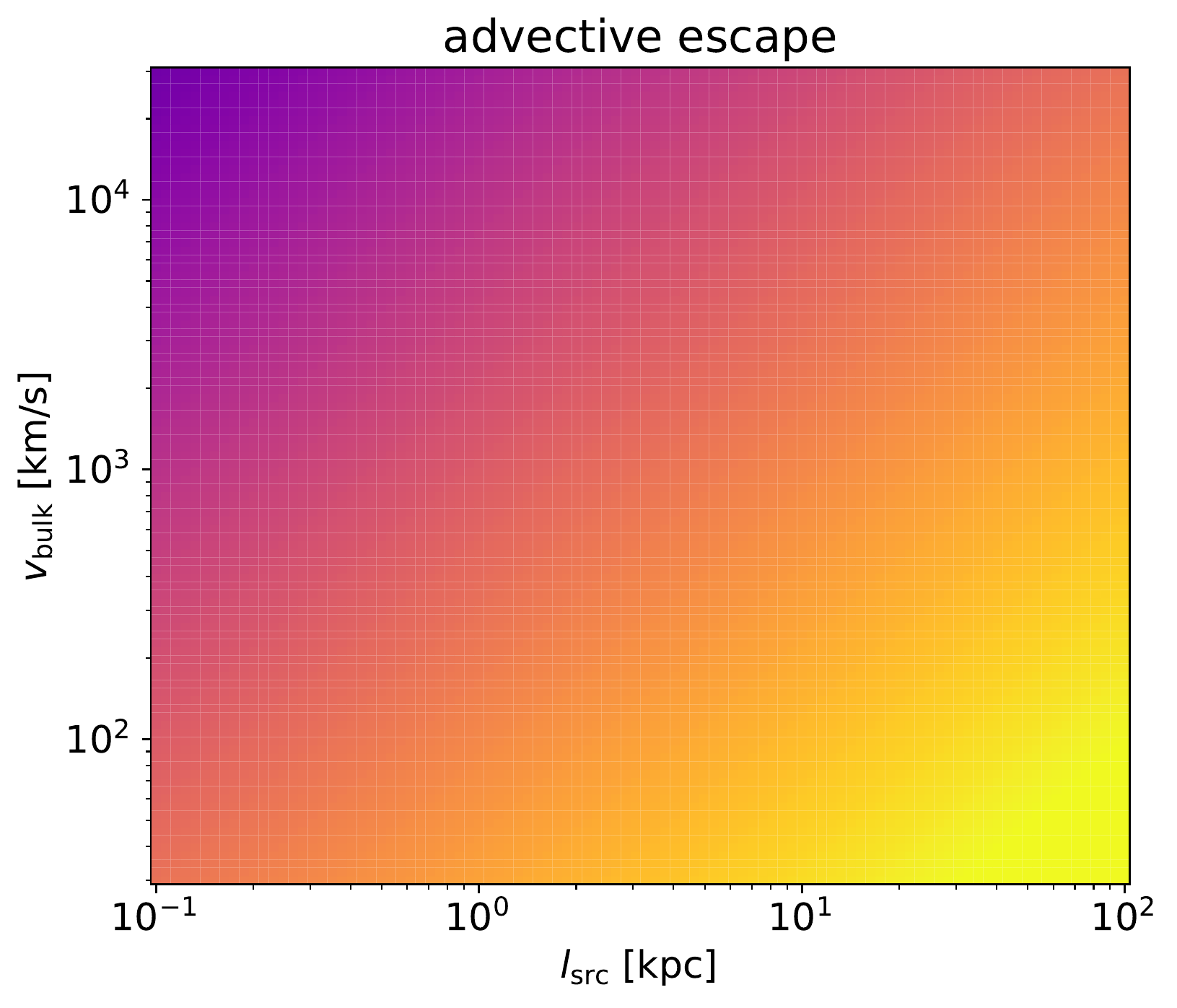}
\includegraphics[width=.465\linewidth]{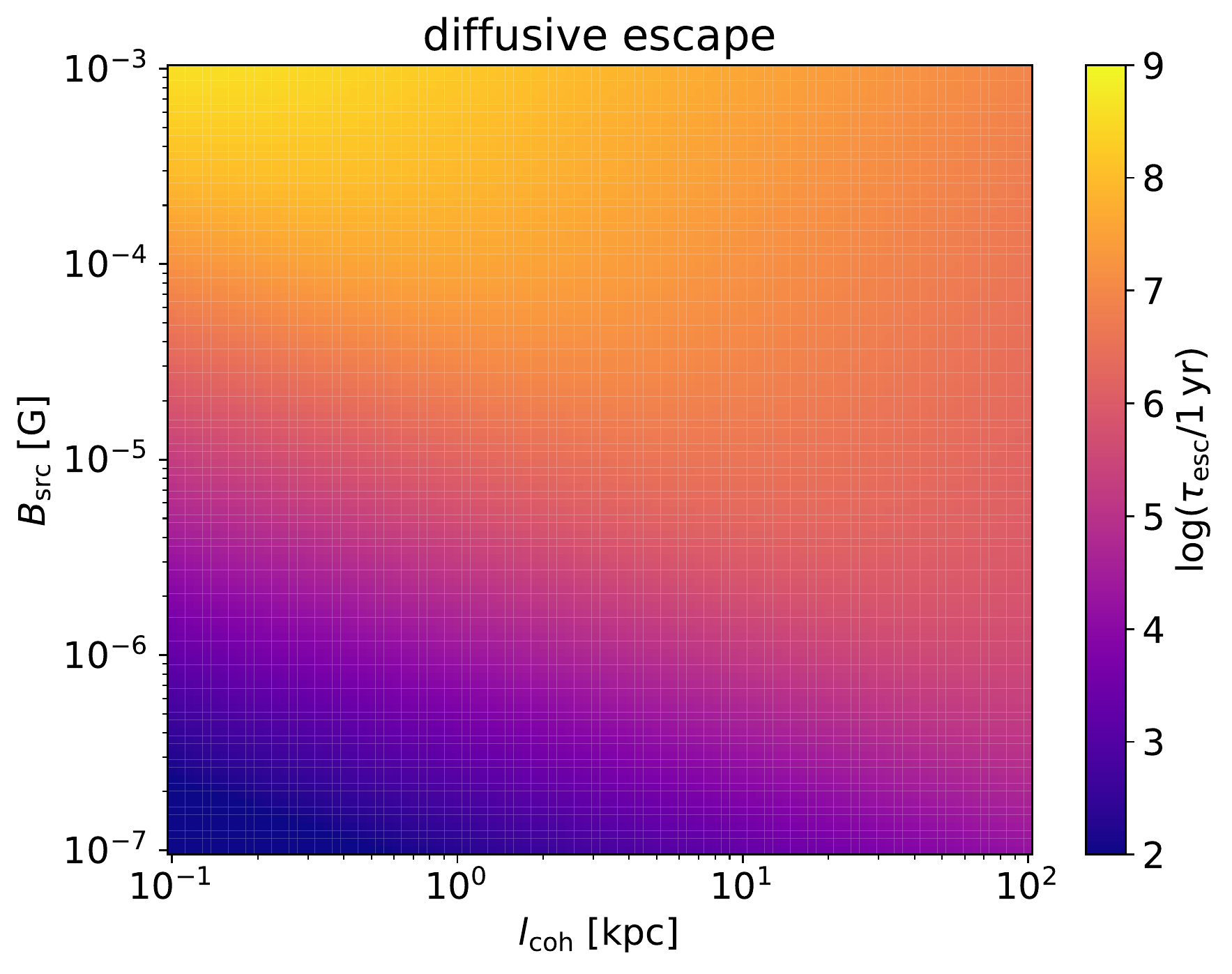}
\caption{The escape time of CRs with a rigidity of 10\,EV in the case of advection (\emph{left}) and Kolmogorov diffusion (\emph{right}) based on Eq.~\ref{eq:escTimes}. In the diffusion case a characteristic size of $l_{\rm src}=100\,\text{kpc}$ is adopted. Note that Kraichnan or Bohm turbulence yields only small modifications of the pattern.}
\label{fig:escTimes}
\end{figure}

While we remain agnostic about the details
of the acceleration process, we have to take into account that UHECRs do not
immediately leave the source region. The details of the escape process are in
general complicated and dependent on the jet dynamics as well as the
spatial position of the accelerator and the magnetic field structure. Apart
from some additional impact by streaming motions, the CR escape from the
source is expected to be dominated by diffusion or advection. If the relevant
bulk flows that drive the particle escape, e.g.\ the back-flowing jet
material, are rather uniform and reach a significant fraction of the speed
of light~\cite[e.g.][]{Reynolds+2002,Matthews+2019}, UHECRs can escape the
source on Myr time scales via advection. But in particular at large
distances from the shock, the bulk flows are typically slower,
i.e.~$v_{\rm bulk}\sim 1000\,\mathrm{km/s}$, and less uniform so that the
UHECR escape is rather driven by diffusion. In addition, the UHECR sources
can be embedded in a galaxy cluster---which is often the case for radio-loud
AGNs---so that in any case the CRs scatter off the turbulent cluster magnetic
field and diffuse out of this structure. In the following, we characterize
the escape from the source region with the characteristic size $l_{\rm src}$ by
the escape time~\cite{MatthewsTaylor2021,Harari+2014}
\be
\tau_{\mathrm{esc}}\simeq\begin{cases}
     9.81~\mathrm{Myr}\left(\frac{l_{\mathrm{src}}}{100~\mathrm{kpc}}\right)\,\left( \frac{v_{\rm bulk}}{10000\,\mathrm{km/s}} \right)^{-1},\,&\text{for advection,}\\
     9.77~\mathrm{Myr}\left( \frac{l_{\mathrm{src}}}{100~\mathrm{kpc}} \right)^2 \left( \frac{l_{\mathrm{coh}}}{1~\mathrm{kpc}} \right)^{-1}  \left[ 4 \left( \frac{R}{R_{\rm c}} \right)^2 + a_{\rm I} \left(\frac{R}{R_{\rm c}}\right) + a_{\rm L} \left( \frac{R}{R_{\rm c}} \right)^{2-m} \right],\,&\text{for diffusion}\,,
    \end{cases}
    \label{eq:escTimes}
\ee
where we account for the quasi-rectilinear propagation at rigidities
significantly higher than the critical rigidity
$R_{\rm c}=9\,(B_{\rm src}/10\,\mu\text{G})\,(l_{\rm coh}/1\,\text{kpc})\,\text{EV}$. Moreover, $B_{\rm src}$ denotes the strength of the turbulent magnetic field
within the source environment and $l_{\rm coh}$ is its coherence length.
Dependent on the spectral index $m$ of the turbulence spectrum, the
coefficients are given by $a_{\rm I}=0.9$ and $a_{\rm L}=0.23$ for Kolmogorov
($m=5/3$) and $a_{\rm I}=0.65$ and $a_{\rm L}=0.42$ for Kraichnan turbulence
($m=3/2$).\footnote{Since the work by Harari et al.\ does not investigate the
  case of Bohm diffusion ($m=1$), we stick to the Kraichnan coefficients
  $a_{\rm I}=0.65$ and $a_{\rm L}=0.42$ in that case.} 
Hence, the CR luminosity that escapes the sources is determined by
\begin{align}
\label{eq:escLcr}
L_{\rm cr}^{\rm esc}(t^\prime)&=\frac{1}{\tau_{\mathrm{esc}}}\int_{-\infty}^{t^\prime}\diff t^{\prime\prime}\,\,L_{\rm cr}(t^{\prime\prime})\,\exp\left(- \frac{(t^\prime-t^{\prime\prime})}{\tau_{\mathrm{esc}}} \right)\nonumber\\
&= \frac{L_0\,\sigma_{\rm act}\, \sqrt{\pi/2}}{\tau_{\mathrm{esc}}\exp\left( - (t_{\rm max}/\sigma_{\rm act})^2/2 \right)}\,
\exp\left( \frac{\sigma_{\rm act}^2-2\tau_{\mathrm{esc}}(t^\prime + d_{\rm src}/c-t_{\rm max})}{2\tau_{\mathrm{esc}}^2}\right)\\
&\qquad\times \left[ 1 - \mathrm{erf}\left( \frac{\sigma_{\rm act}^2-\tau_{\mathrm{esc}}(t^\prime+d_{\rm src}/c-t_{\rm max})}{\sqrt{2}\sigma_{\rm act}\tau_{\mathrm{esc}}} \right) \right]\,, \nonumber
\end{align}
where $\mathrm{erf}(x)$ denotes the error function. It can be shown that
$L_{\rm cr}^{\rm esc}(t^\prime) \rightarrow L_{\rm cr}(t^\prime)$ for
$\tau_{\mathrm{esc}} \rightarrow 0$. In the opposite limit,
$\tau_{\mathrm{esc}}\gg t_{\rm act}$, the CR contribution of the present,
individual sources becomes negligible. Hence, a present source with a
life-time of a few Myr and a characteristic size of about 100\,\text{kpc} can
only provide a significant contribution of 10\,EV CRs, if either a bulk flow
with $v_{\rm bulk}\gtrsim 10000\,\text{km/s}$ or a magnetic field with
$B_{\rm src}\lesssim 10\,\mu\text{G}$ is present, cf.\ with
Fig.~\ref{fig:escTimes}.
\begin{figure}[htbp]
\centering
\includegraphics[width=.45\linewidth]{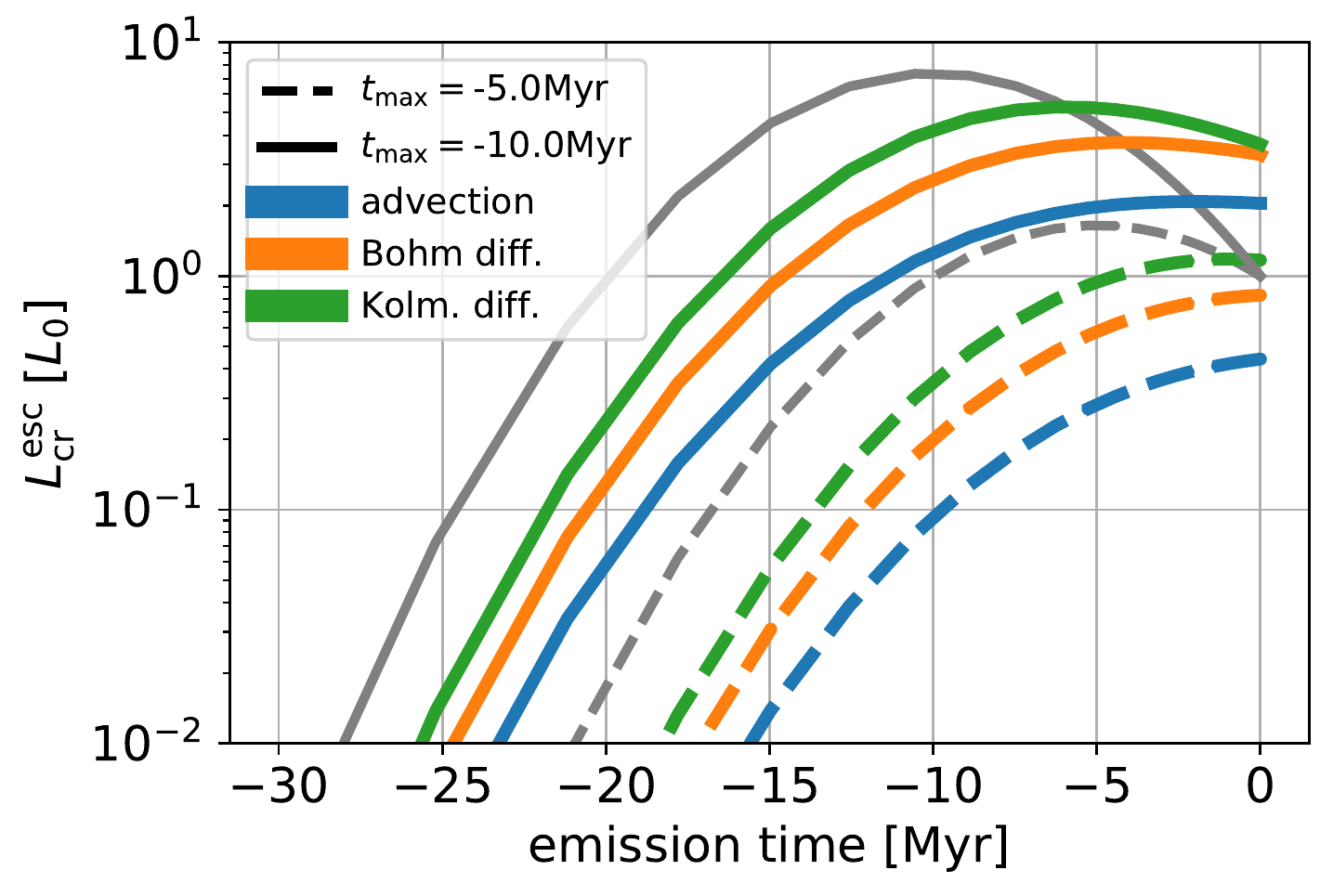}
\includegraphics[width=.45\linewidth]{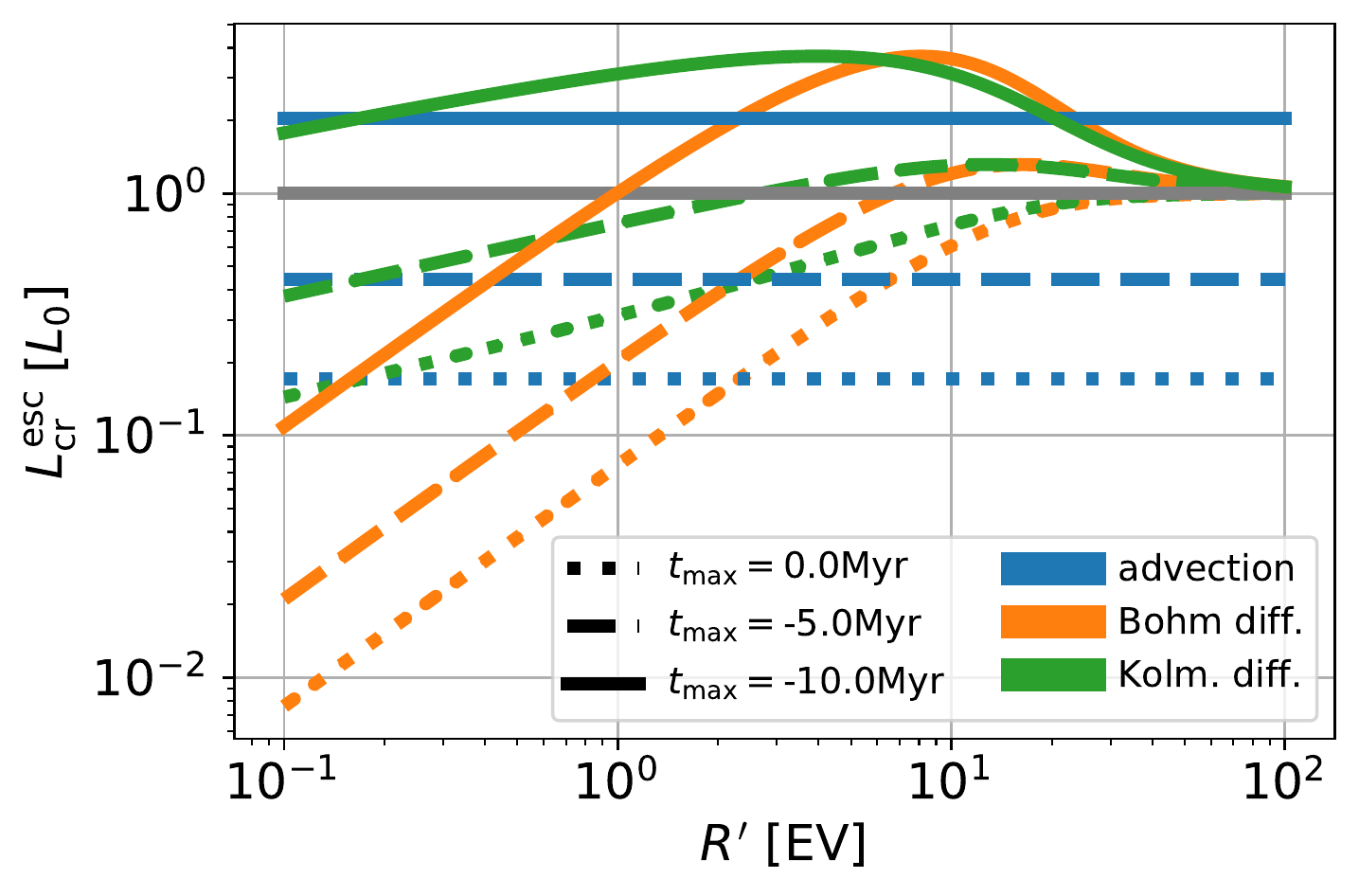}
\caption{The escaping luminosity $L_{\rm cr}^{\rm esc}$ in the case of advection (blue), Bohm diffusion (orange) and Kolmogorov diffusion (green), respectively,
  $l_{\rm src}=100~\mathrm{kpc}$, $B_{\rm src}=10~\mu\mathrm{G}$, $l_{\rm coh}=10~\mathrm{kpc}$, $v_{\rm bulk}=3000~\mathrm{km/s}$ and an activity time of $t_{\rm act}=15\,\mathrm{Myr}$. The grey lines indicate the luminosity in case of an instantaneous escape from the source.\newline
\emph{Left:} The evolution of the source luminosity dependent on the emission time ($-t^\prime-d_{\rm src}/c$) for different $t_{\rm max}$ values for a CR rigidity $\Rin=1\,\mathrm{EV}$.
\emph{Right:} The spectral behaviour of the present ($-t^\prime=d_{\rm src}/c$) source luminosity for different $t_{\rm max}$ values.}
\label{fig:escLcr}
\end{figure}
For a given initial CR luminosity $L_{0}$ at the present (with respect to
the light travel time distance)  time, it is illustrated in
Fig.~\ref{fig:escLcr} that the past maximal value of the CR luminosity
$L_{\rm cr}$ can differ by orders of magnitude, depending on the given
activity time $t_{\rm act}$ as well as the time $t_{\rm max}$ of maximal
emission. In the limit $t_{\rm act}\gg |t_{\rm max}|$, this difference obviously
vanishes. When we account for  advective or diffusive escape, respectively,
the increased past source activity gets delayed so that the present escaping
luminosity $L_{\rm cr}^{\rm esc}(-d_{\rm src}/c)$ becomes typically maximal if
$t_{\rm max}\sim -\tau_{\mathrm{esc}}$. In that case the luminosity
$L_{\rm cr}^{\rm esc}(-d_{\rm src}/c)$ can also exceed $L_{0}$, which does not
account for any CR propagation effects since this quantity is typically derived from the present luminosity in some particular band of the electromagnetic spectrum. Since the diffusive escape of the CRs
from the source region depends on the particles' rigidity, the resulting
emissivity does so too: The total number of escaping CRs typically increases
with increasing rigidity leading to a significant hardening of the escaping
energy spectrum, as can be seen in the right panel of Fig.~\ref{fig:escLcr}.
Note that this is not the case if the CRs escape by advection.
However, the subsequent propagation through the extragalactic magnetic field
can still lead to a hardening of the CR energy spectrum at Earth if the sources
have a finite life-time. Moreover, for the limiting case of Bohm diffusion
the hardening of the spectrum only yields a spectral index of $\alpha-1$ at low
rigidities. Considering a spectral index of $\alpha=2$ at the acceleration site
this hardening alone is clearly not sufficient to reproduce the composition
data. 

Although, it is most reasonable to adopt a finite life-time of the considered
UHECR sources, this also introduces the difficulty of modelling those sources
that are currently inactive with respect to their electromagnetic emission
but still contribute UHECRs due to their past activity.  At least on large
scales, we can use a redshift dependent luminosity function to account for
those additional sources.

\subsection{Continuous CR sources}
\label{sec:csf}

Based on the radio luminosity function of low- and high-luminosity radio
sources from Ref.~\cite{Willott+2001} which provides the number per volume
and luminosity $\diff N /\diff V\,\diff L_{\rm cr}$ of radio sources, we
use the so-called continuous source function (CSF), 
\begin{equation}
  \Psi_{i}(\Rin,\,z) \equiv
  \frac{\diff N_{\rm cr}(\Zin_i) }{ \diff V \mathrm{d}\Rin\,\diff t} = 
\int \left( \frac{\diff N}{\diff \Rin}\right)_{\rm sim}\,w_{\rm R}\big(\Rin,\hat R(L_{\rm cr})\big)\,\frac{\diff N }{
\diff V\,\diff L_{\rm cr}}\,\diff L_{\rm cr}\,,
\label{CRsourceRateDensity}
\end{equation}
that has already been introduced in previous works, e.g.\
Refs.~\cite{Eichmann2019,Eichmann2019ICRC}, to obtain the time (redshift)
dependent CR emissivity arising from the background population of sources on
large scales. Here, we suppose that the CSF is homogeneous in space,
but even for a discrete distribution of sources we can still suppose that
the universe is homogeneously filled with CRs if they propagate at least
the average source distance~\cite{AloisioBerezinsky2004}.
Moreover, $\left( \diff N/\diff \Rin \right)_{\rm sim}$ denotes the initial
rigidity distribution of UHECRs that have been simulated in 1D with the
UHECR propagation code CRPropa3~\cite{CRPropa3_2016, CRPropa3.1_2017, CRPropa3.2_2022}.
The weight $w_R$ has been introduced to modify the initial spectrum
afterwards (see the Appendix~\ref{App:IndSrc} for more details). 
Then the isotropic UHECR intensity is generally given by 
\begin{align}
  J_{\rm csf}(E + \Delta E,\,t_{\rm obs})&=\frac{c}{4\pi} \sum_{\epsilon,i,j}\Delta z_j\,\,\left| \frac{\diff t}{\diff z}\, \right|\,\frac{\Rin(\epsilon)\,\Psi_{i}(\Rin(\epsilon),\,z_j)}{\Delta E}\,,
                                           \label{eq:Jcsf}
\end{align}
where we sum over all CR particles species $i$, all energies $\epsilon\in[E, E + \Delta E]$ within the energy bin of interest,
and all source distances $z_j\in [0,\,2]$. Since the
source sample considered by us consists of only a few sources, it is appropriate to extend the CSF up to $z=0$. As shown in the Appendix~\ref{App:CSF}, the previous equation
can be simplified for a  1D distribution of sources uniform in light travel
distance. In a similar manner, but based on 3D simulations, also the flux
$J(E + \Delta E,\,t_{\rm obs})$ from individual sources with the temporal
CR luminosity given by Eq.~(\ref{eq:escLcr}) can be treated; for details
see Appendix~\ref{App:IndSrc}.

\section{Conditions for the dominance of individual sources}
\label{sec:ind-to-csf}

The contribution~(\ref{eq:Jcsf})  from the CSF to the UHECR intensity is
much softer than the one from individual sources with a finite life-time,
even if their source spectral indices are comparable. This difference is a
result of the magnetic horizon~\cite{Parizot:2004wh,Berezinsky:2005fa} which
prevents that CRs with low rigidity from individual sources do reach the
observer if the source life-time is too short. We expect that there
exists a critical energy for a given source luminosity $L_0$ where
the contribution from individual sources will start to dominate over the
one from the CSF. Based on the observed spectral features, the hardening above the ankle at about 5\,EeV could indicate this transition. In general, the value of this critical energy depends on a
multitude of parameters, such as the properties of the extragalactic magnetic
field, the UHECR composition, the escape from the source
environment as well as the temporal features of the activity phase, and
the source distance. In addition, the CSF contribution is strongly dependent
on the assumed radio luminosity function as well as the radio-jet power
correlation. 

\begin{figure}[htbp]
\centering
\includegraphics[width=.45\linewidth]{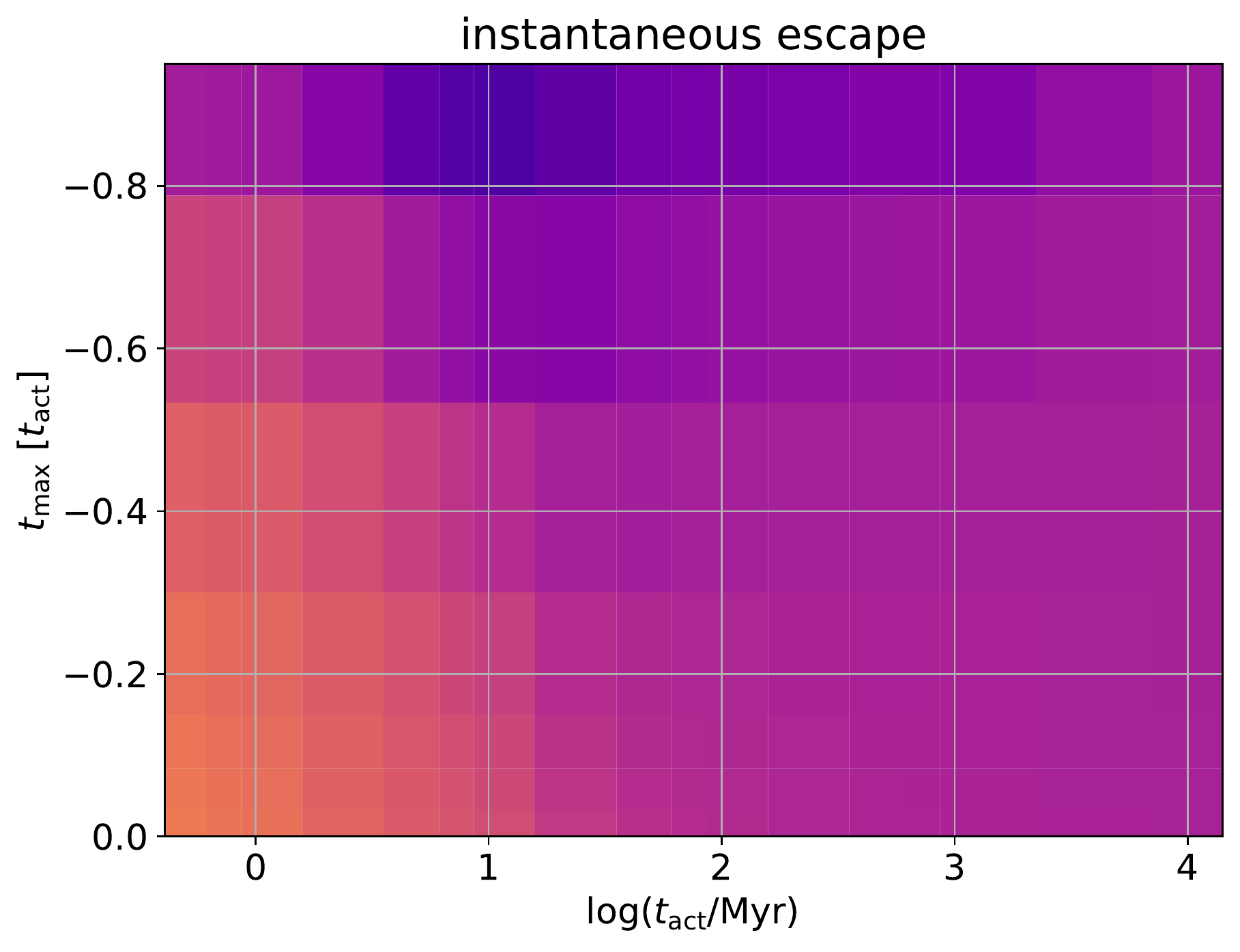}
\includegraphics[width=.45\linewidth]{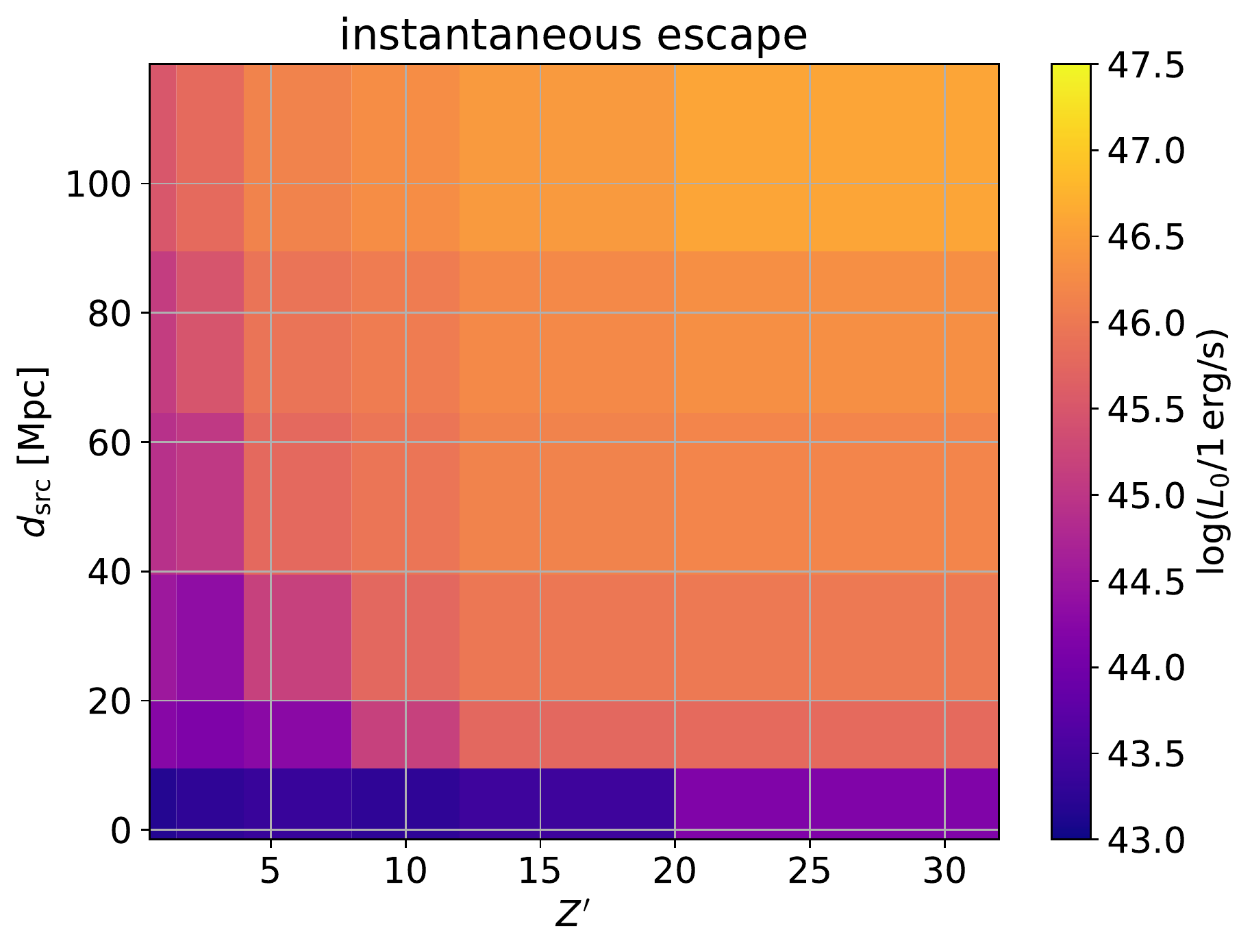}
\includegraphics[width=.45\linewidth]{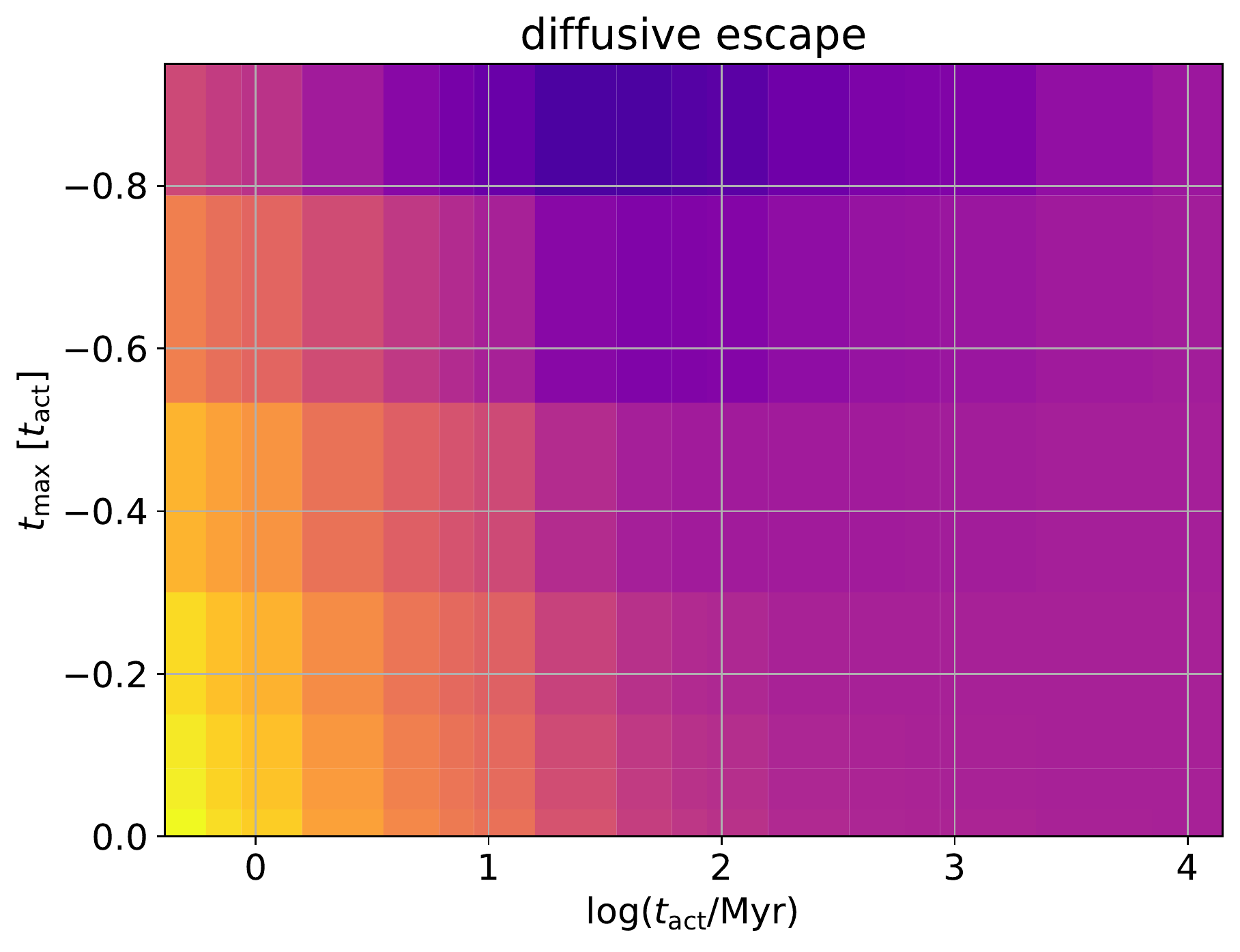}
\includegraphics[width=.45\linewidth]{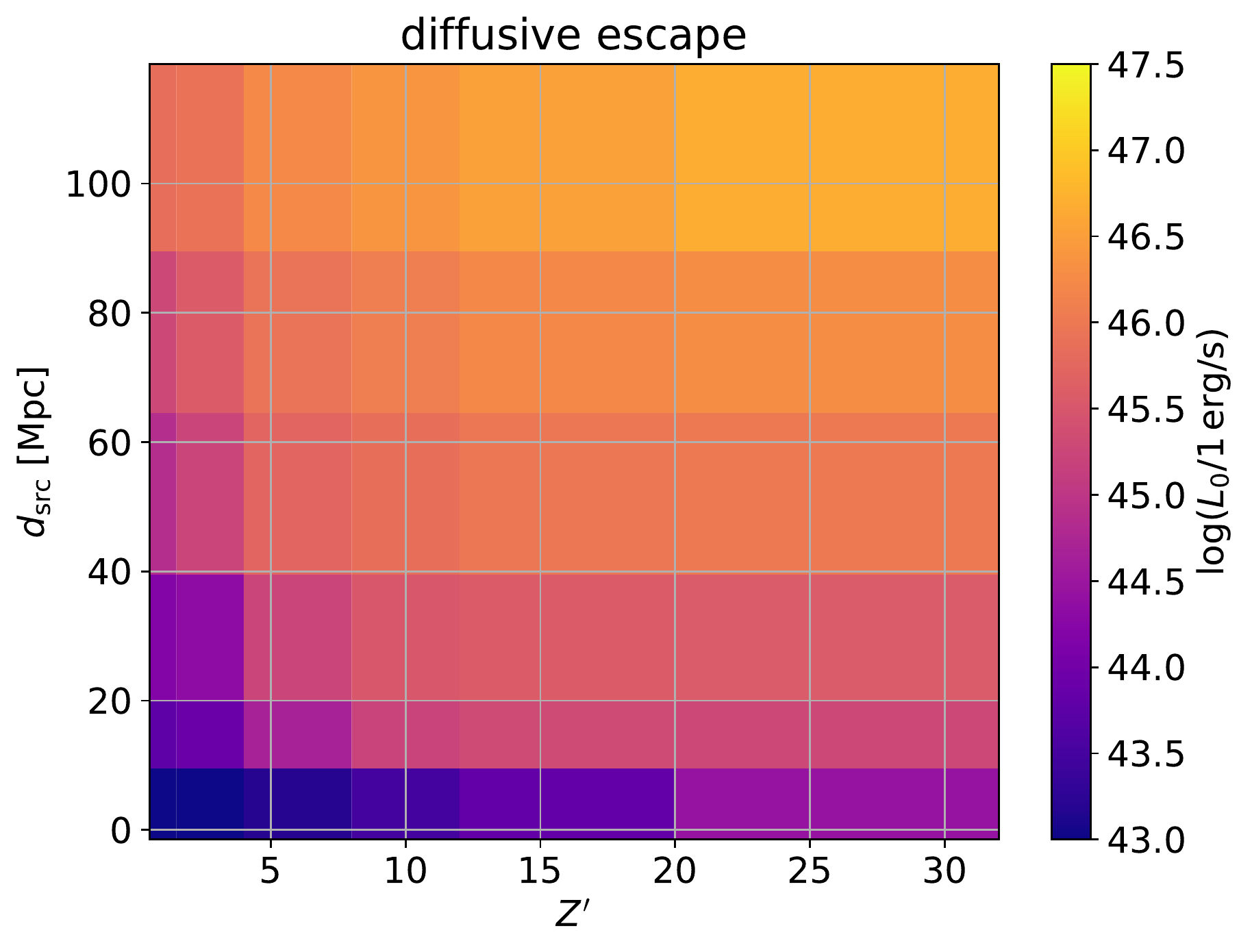}
\includegraphics[width=.45\linewidth]{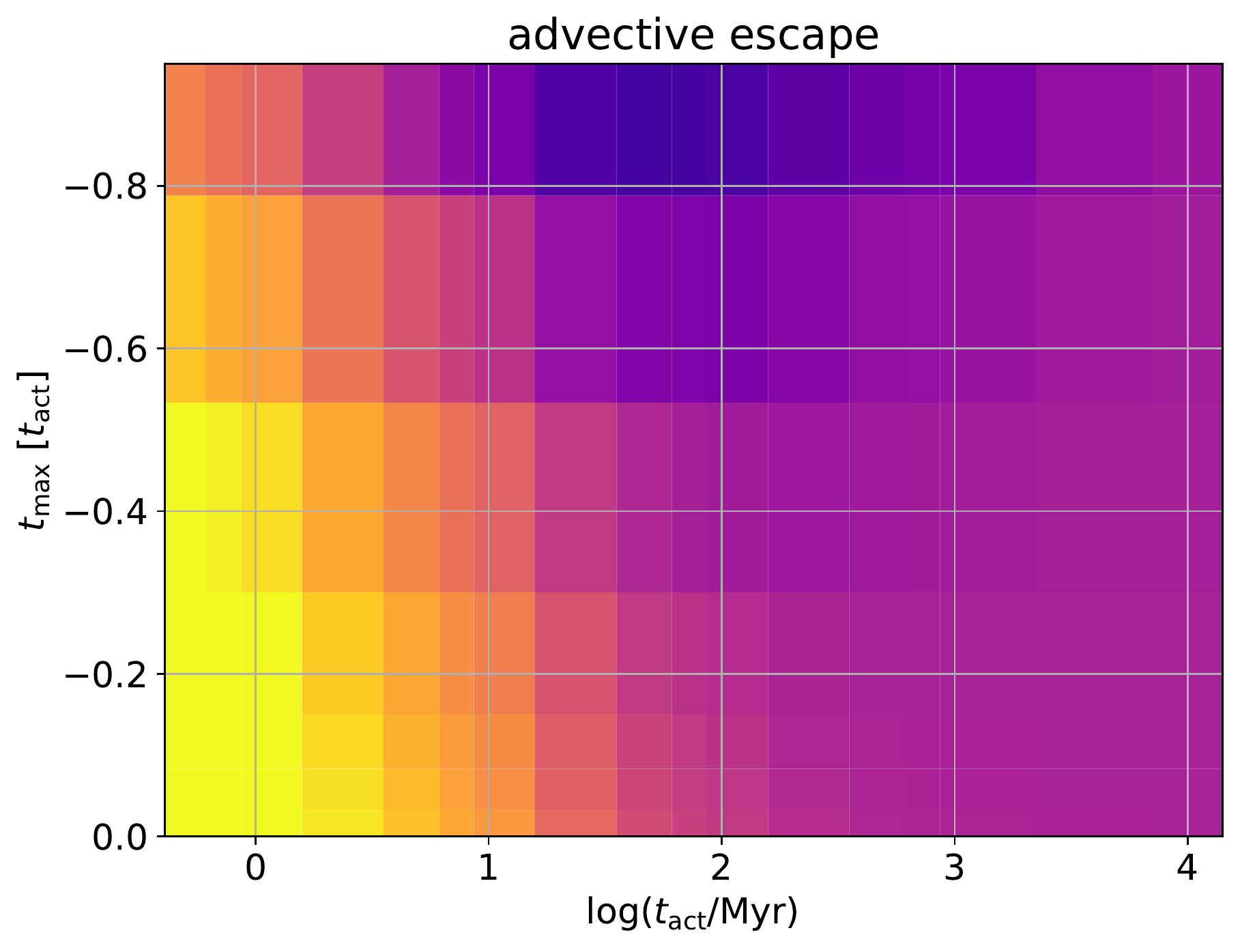}
\includegraphics[width=.45\linewidth]{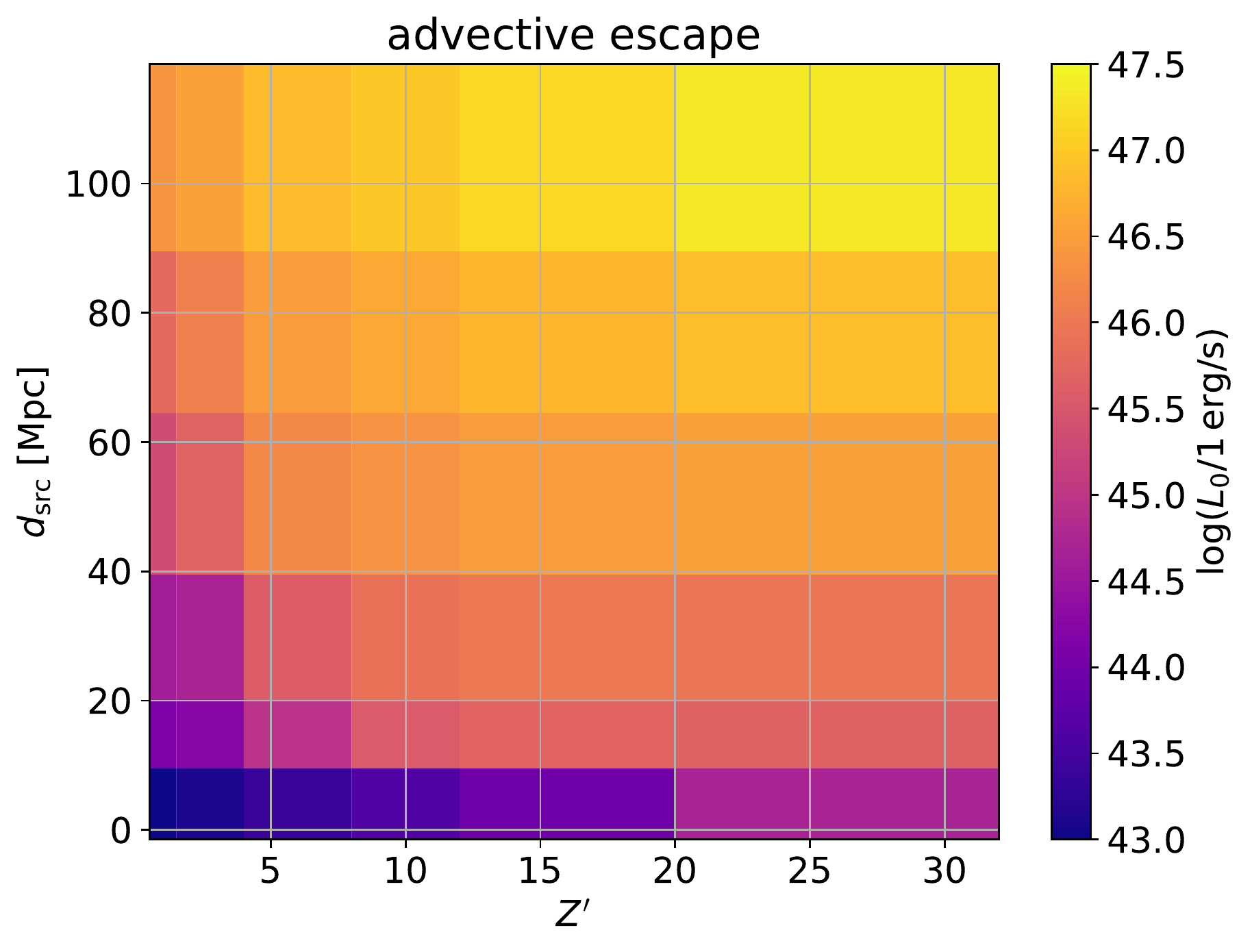}
\caption{The necessary CR luminosity $L_{0}$ of an individual source for different escape scenarios (instantaneous--\emph{upper panel}; Kolmogorov diffusion--\emph{middle panel}; advection--\emph{lower panel}) to dominate the observed UHECR spectrum above the ankle (at $E=5\,\text{EeV}$). On the left hand side, it is supposed that the ejected chemical composition is dominated by carbon and the source is at a distance $d_{\rm src}=15\,\mathrm{Mpc}$. On the right hand side, we check the dependence on those assumptions by keeping the characteristics of the activity time fixed ($t_{\rm max}=-4\,\mathrm{Myr}$ and $t_{\rm act}=10\,\mathrm{Myr}$). The adopted escape parameters for the diffusive and advective scenario, respectively, are $l_{\rm src}=100$\,kpc, $B_{\rm src}=10\,\mu\mathrm{G}$ and $v_{\rm bulk}=3000\,\mathrm{km/s}$.}
\label{fig:ind-csf-rel}
\end{figure}

Despite this rather complicated dependencies, we would like to determine
the typical luminosity $L_0$ of an individual source at a given distance
$d_{\rm src}$ required to dominate the observed UHECR flux above the ankle.
In order to reduce the parameter space, we consider only cases where the
flux is dominated by a single nucleus type,
as mixed scenarios can be interpolated from that. Furthermore, we suppose a
generic maximal rigidity of
$\hat{R}=50\,g_{\rm acc}/\sqrt{1-g_{\rm m}}$\,EV for the considered
source. This ensures that this source can in principle explain the observed
UHECR flux up to the highest energies. Moreover, we adopt $g_{\rm m}=4/7$ and
$g_{\rm acc}=1$ both for the individual sources as well as for the CSF.
Since both  contributions depend on these parameters in a similar way, the
following results depend only weakly on these values as long as the cutoff
is well above the ankle.
Note that for these values the UHECR flux contribution from the CSF is about
a factor 2--10 smaller (dependent on the assumed composition) than the flux
at the ankle. Hence, we ensure that the total flux prediction by individual
sources and the CSF has also the right order of magnitude.

In Fig.~\ref{fig:ind-csf-rel}, we show the necessary CR luminosity $L_{0}$ of
an individual source  to dominate the observed UHECR flux above $E=5\,$EeV.
All panels are for a purely turbulent EGMF with a RMS field strength of 1\,nG,
a Kolmogorov spectrum and a coherence length of $0.2\,$Mpc using
wave modes between 60 and 800\,kpc.
For these parameters, CR propagation takes predominantly place in the
small-angle scattering regime. Three
different escape scenarios are compared: instantaneous,
advective and diffusive escape. For the latter we adopt Kolmogorov diffusion,
although Bohm diffusion that is caused by, e.g., the Bell instability~\cite{Bell2004}
could be present at the acceleration site. But since its size is small compared
to the whole source extension of about a few\:$\times 100\,\text{kpc}$, we
expect that the diffusive escape from the source environment is rather
dominated by the Kolmogorov turbulence on large scales. 
On the left hand side of Fig.~\ref{fig:ind-csf-rel}, the UHECR flux consists of carbon and
the source is at a distance $d_{\rm src}=15\,\mathrm{Mpc}$. The contribution of
this source is strongest, i.e.\ the necessary $L_0$ value is the smallest, if
the source is close to its final stage, $t_{\rm max}\sim -t_{\rm act}$. In contrast, young sources with an increasing CR luminosity evolution (i.e.\ $t_{\rm max}>0$) need an extraordinarily high present luminosity to exceed the CSF contribution and to explain the observed flux above the ankle. A larger activity time decreases the necessary $L_0$, since then CRs with a
larger propagation delay introduced by the extragalactic magnetic
field as well as the escape process can still reach the observer.
Before for large activity times the influence of $t_{\rm max}$ vanishes---as
all additional CRs have already reached the observer in the past---there
is a minimal CR luminosity reached. Here, the time $t_{\rm max}$ of maximal
CR emissivity corresponds to the necessary propagation time of the majority
of CRs at that energy. 

In addition to the obvious decrease of $L_0$ with decreasing source distance,
we also notice that the source composition has a significant influence:
This effect is visible on the right hand side of Fig.~\ref{fig:ind-csf-rel}, where we check the dependence
of $L_0$ on the composition and the distance keeping the activity time fixed,
choosing $t_{\rm max}=-4\,\mathrm{Myr}$ and $t_{\rm act}=10\,\mathrm{Myr}$.
A light composition yields a lower value of $L_0$ than a heavy one, which is
predominantly due to the decreasing number of sources in the CFS contribution
that yield energies above the ankle. Hence, $J_{\rm csf}$ increases with
increasing $Z^\prime$ despite $J_{\rm csf}\propto 1/\bar{Z}$, where the average initial charge number $\bar{Z}$ increases (for definition of $\bar{Z}$ see \ref{App:IndSrc}). In contrast, the flux of
an individual source decreases with increasing $\bar{Z}$---especially if it is close-by so that propagation effects are small---according to the flux
normalization~(\ref{eq:weight_ind}). Consequently, distant sources
($d_{\rm src}\sim 100\,\text{Mpc}$) typically need an extraordinarily high CR
luminosity $L_0$ in order to dominate the UHECR flux, in particular if the
composition is dominated by heavy elements.

Comparing the different escape scenarios, there are only minor differences
with respect to the general parameter dependence. However, the additional
delay by diffusive or advective escape generally requires higher $L_0$ values
at small activity times, as less CRs manage to reach the observer in the
given time, and shifts the minimal $L_0$ value towards larger $t_{\rm act}$. 
For all scenarios, we need at least $L_0\sim 10^{44}\,\mathrm{erg/s}$ to
have an individual source dominating over the UHECR contribution from the
large scale population above the ankle. However, we note that the actual
transition energy has a significant impact on these results. 
Individual sources with $t_{\rm act}\lesssim 10\,\text{Myr}$ can only provide the
necessary CR power to dominate the CR spectrum above the ankle, if these source are rather old (i.e.\ $t_{\rm max}\sim -t_{\rm act}$) and located at a distance of only a few\:$\times 10\,\text{Mpc}$. 

\begin{figure}[htbp]
\centering
\includegraphics[width=.55\linewidth]{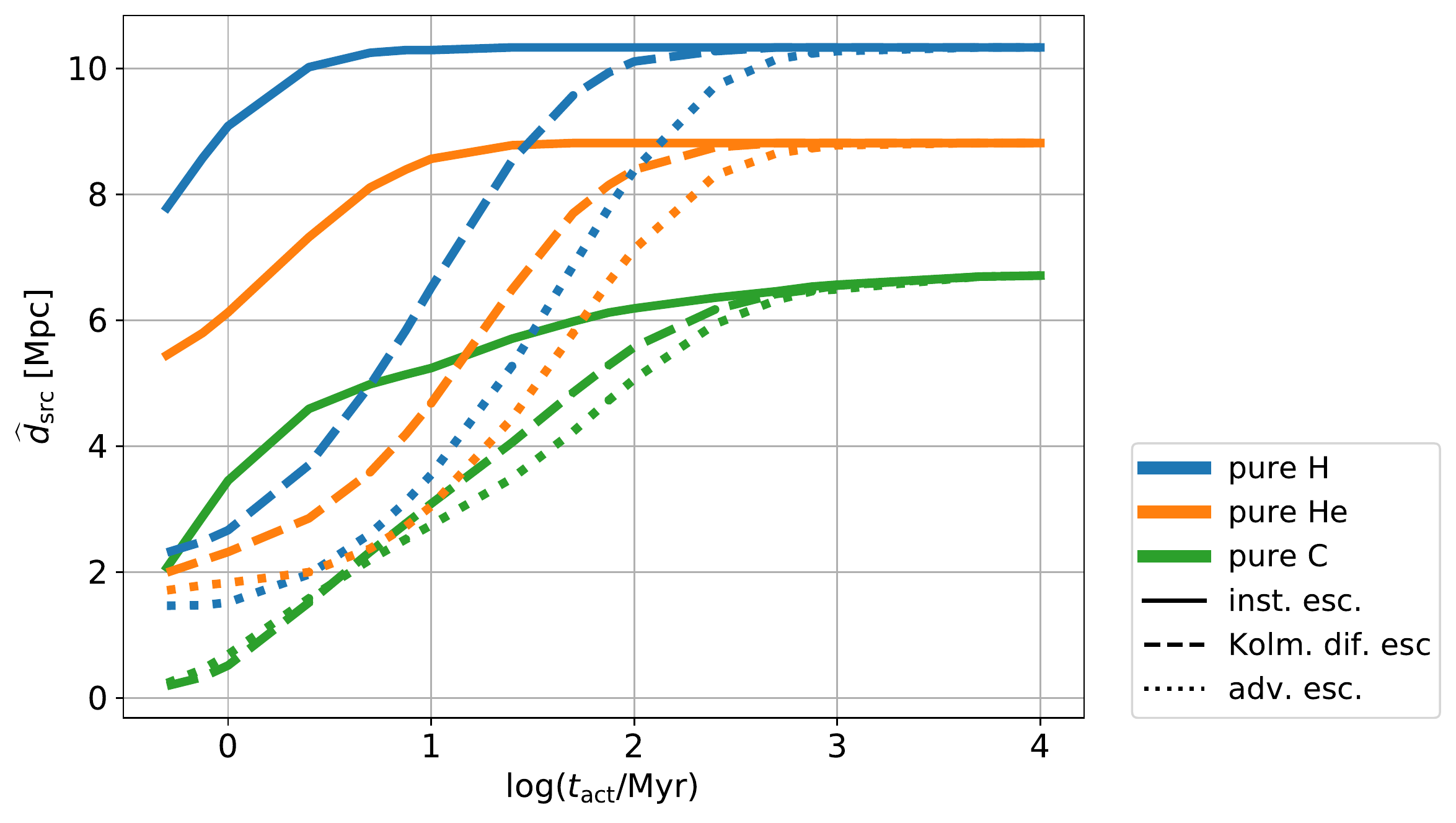}
\includegraphics[width=.41\linewidth]{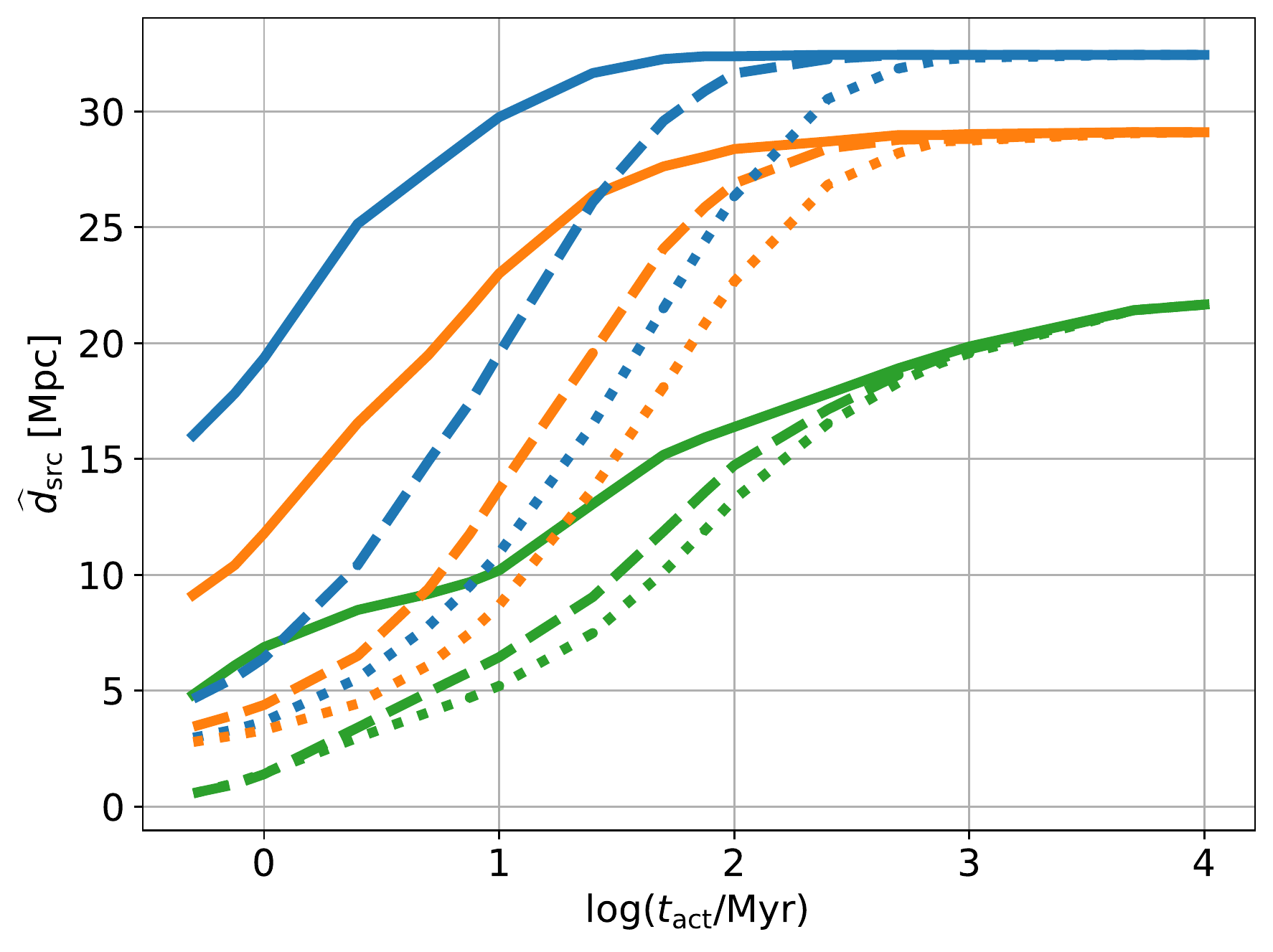}
\caption{The maximal distance $\hat{d}_{\rm src}$ of an individual CR source (to dominate the UHECR flux above the ankle) with a CR luminosity of $L_0=10^{44}\,\text{erg s}^{-1}$ (\emph{left}) and $L_0=10^{45}\,\text{erg s}^{-1}$ (\emph{right}), respectively, dependent on its activity time. It is adopted that this source is currently observed at its maximal emission stage and emits only a single type of nucleus as indicated by the colours.}
\label{fig:dmax-tact-rel}
\end{figure}

Based on an individual source with a given present CR luminosity,
Fig.~\ref{fig:dmax-tact-rel} shows its maximal distance $\hat{d}_{\rm src}$
dependent on the activity time. 
Here, we consider a middle-aged source---i.e.\ it is currently (under consideration of the light travel time) in its maximal emission stage. 
For an optimal choice of $t_{\rm max}$, which is typically
$t_{\rm max}\sim - t_{\rm act}$ (as in that case the source emissivity has been
much higher in the past), the maximal distance  can significantly increase.
For instance, for $t_{\rm max}= -0.7\, t_{\rm act}$ we obtain for certain
activity times a maximal distance  that is about twice as large as
the one shown in Fig.~\ref{fig:dmax-tact-rel}. 
Independent of $t_{\rm max}$ all curves necessarily converge towards large
$t_{\rm act}$, approaching the steady state scenario. This limit yields also
an upper
limit of the source distance dependent on its CR composition and luminosity,
if the source's activity has not been significantly higher in the past. Thus,
Fig.~\ref{fig:dmax-tact-rel} exposes clearly that the limited life-time of
the sources constraints the potential sources to a quite small number of
close-by objects. Hence, in the case of a strong, turbulent EGMF (of $1\,\text{nG}$ rms strength) and a present CR luminosity of $L_0=10^{45}\,\text{erg s}^{-1}$, there are just a very few local radio galaxies, such as Centaurus~A, Virgo~A, Fornax~A and 3C~270, that manage to exceed the CSF contribution at the ankle for a heavy initial composition and $t_{\rm act}\lesssim 1\,\text{Gyr}$.
Larger local samples might be still possible for an appropriate choice of
the parameters, such as a different acceleration efficiency (i.e.\ different
values of $g_{\rm m}$ and $g_{\rm acc}$) of the individual local sources and
the CSF or a significantly higher CR luminosity of the local source(s) in the
past (i.e.\ $t_{\rm max}\sim -t_{\rm act}$).

These results already show that the actual parameter space that allows a
dominant CR flux by individual local sources of a given CR luminosity is
much more limited than one may have expected in the first place. Moreover, it still
needs to be quantified, if a given set of a few local CR sources is actually
able to explain the observed UHECR spectrum and its mass composition.

\section{Constraints from composition data}
\label{sec:compConstr}

\begin{figure}[htbp]
\centering
\includegraphics[width=.45\linewidth]{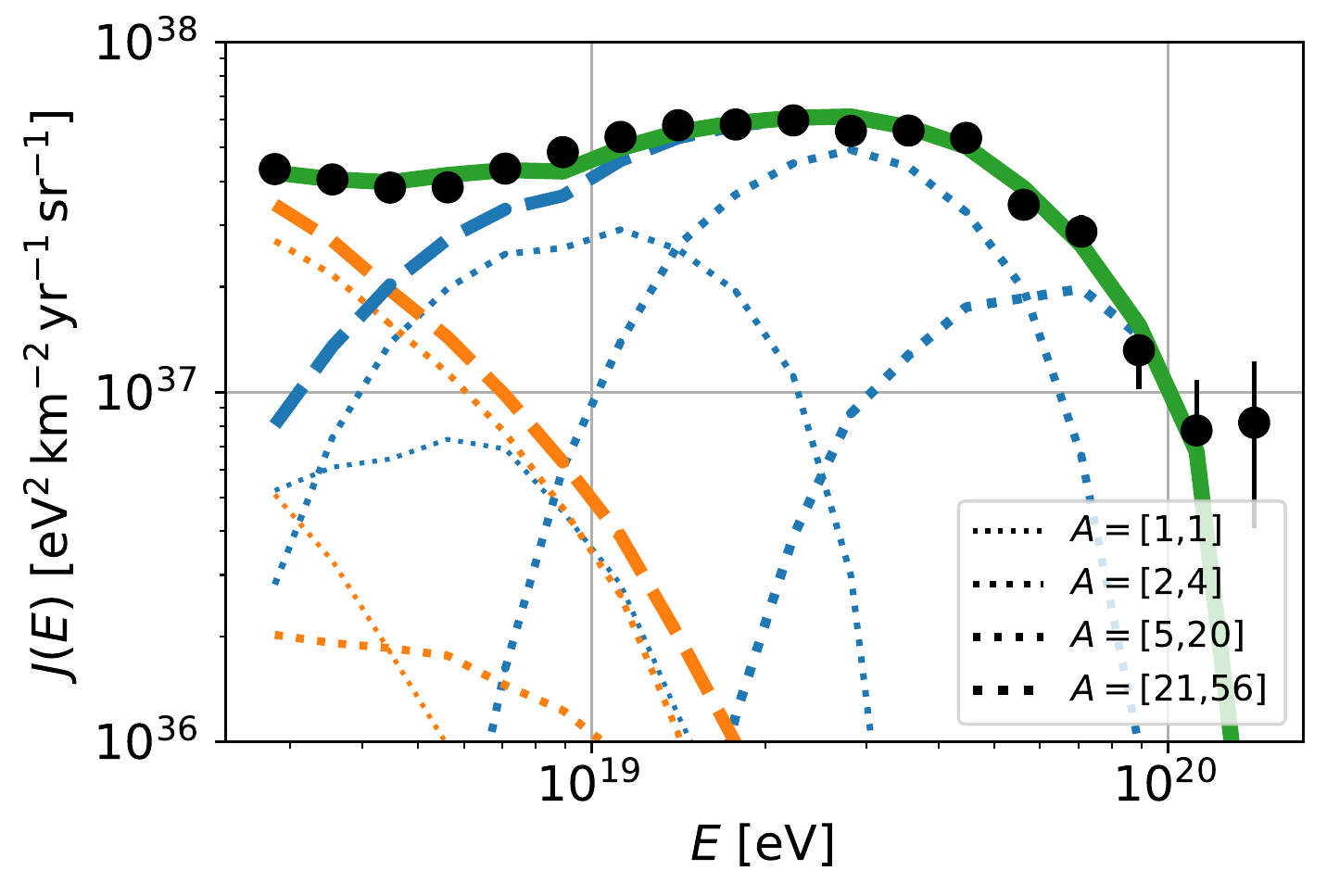}
\includegraphics[width=.45\linewidth]{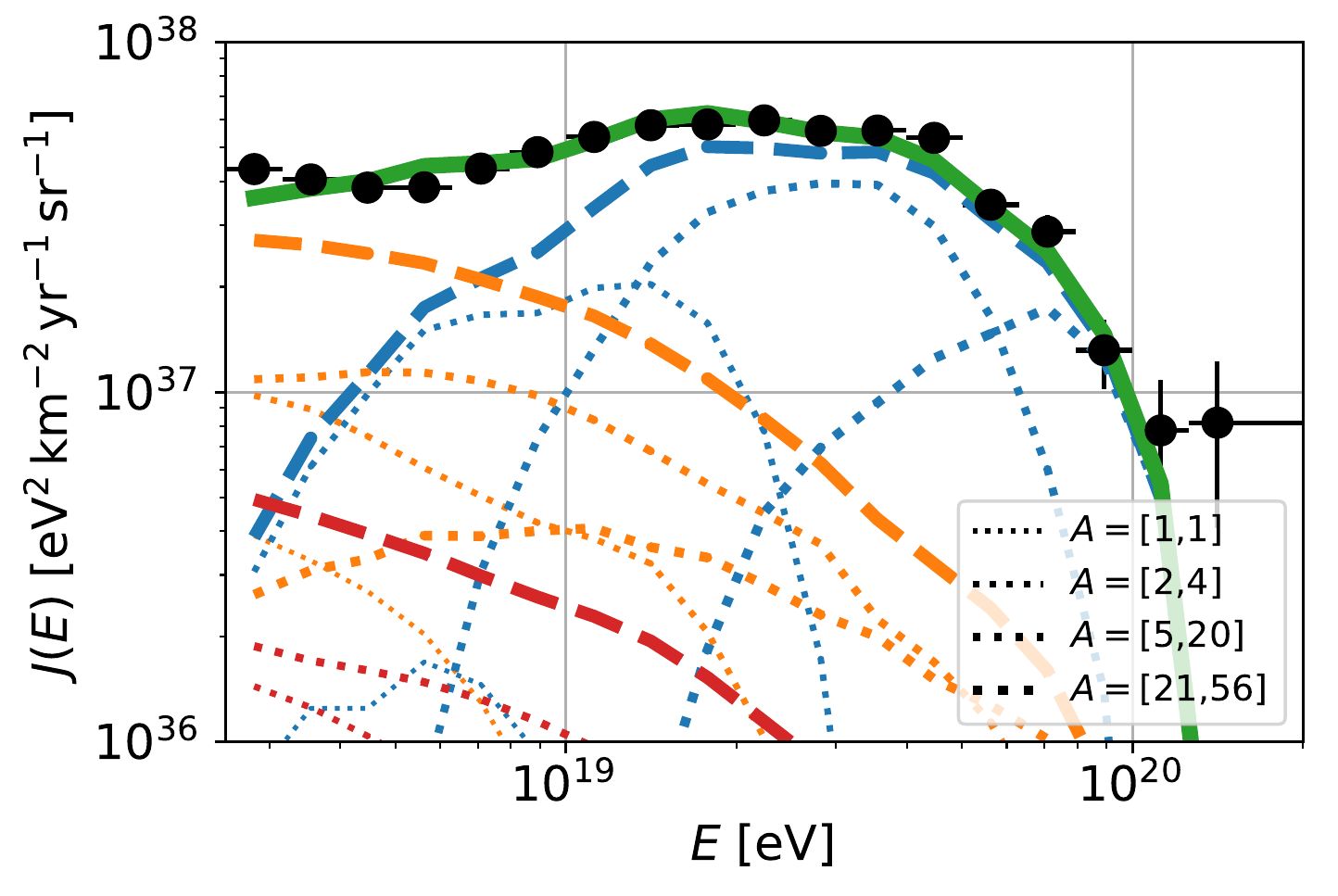}
\includegraphics[width=.45\linewidth]{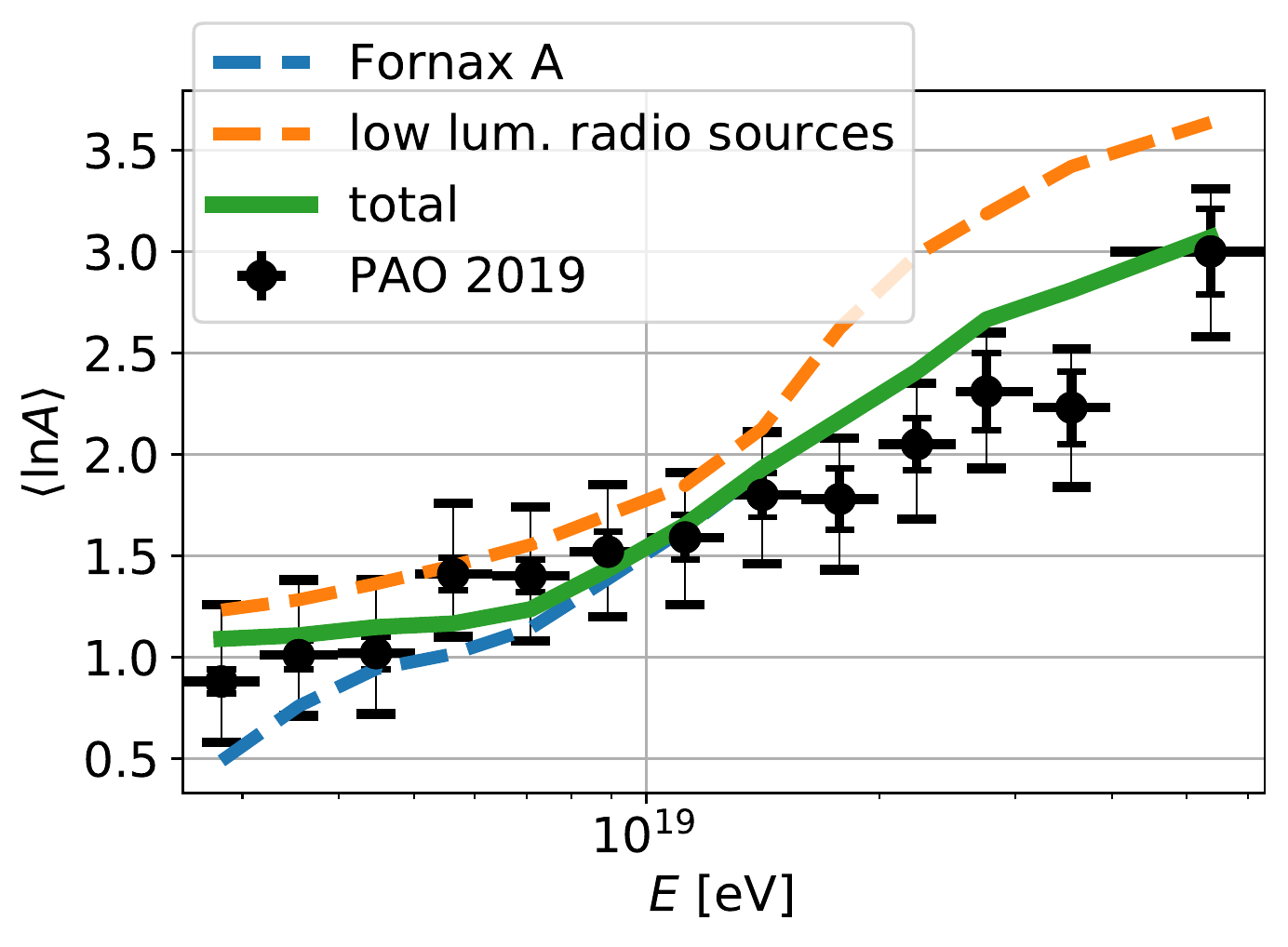}
\includegraphics[width=.45\linewidth]{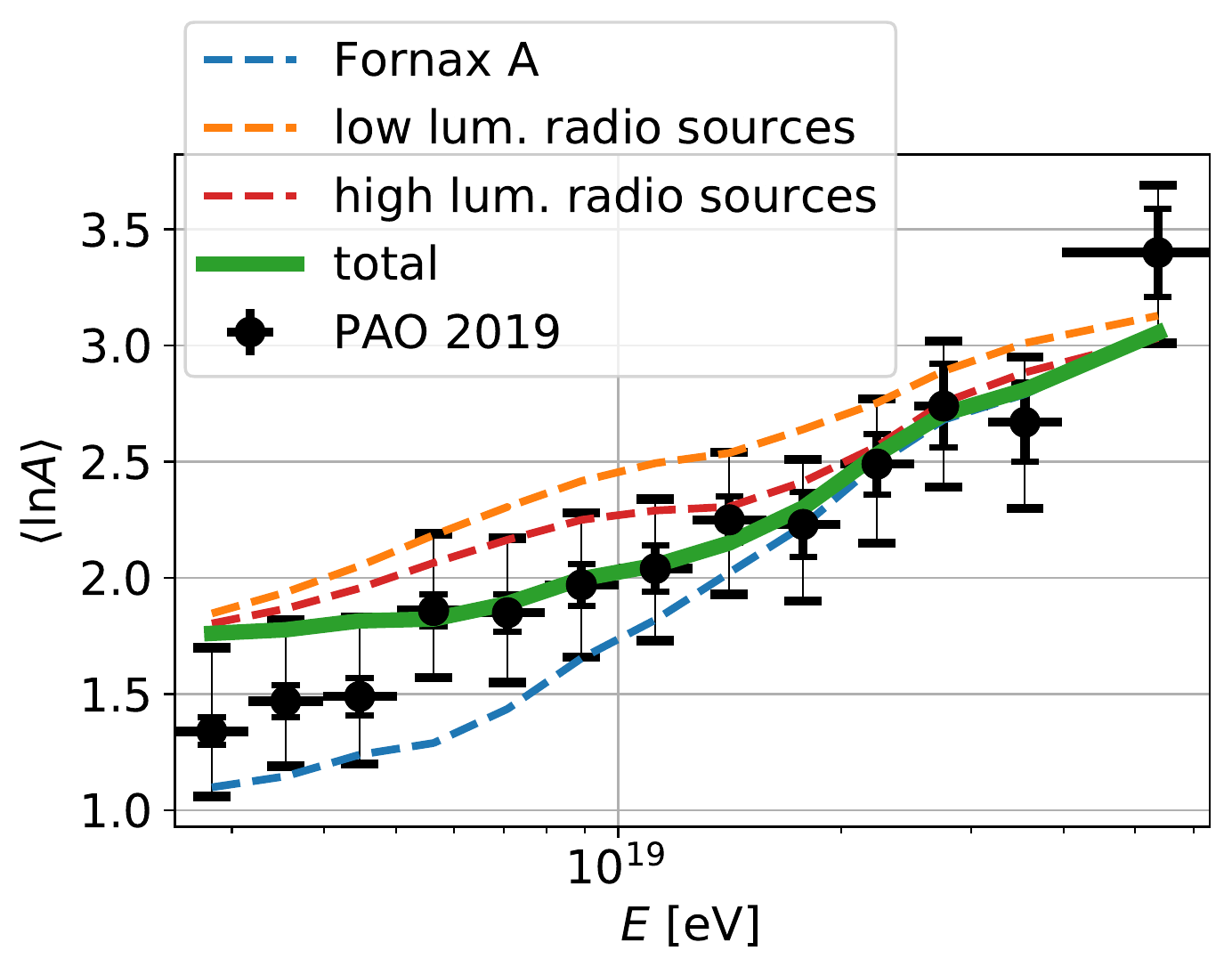}
\includegraphics[width=.45\linewidth]{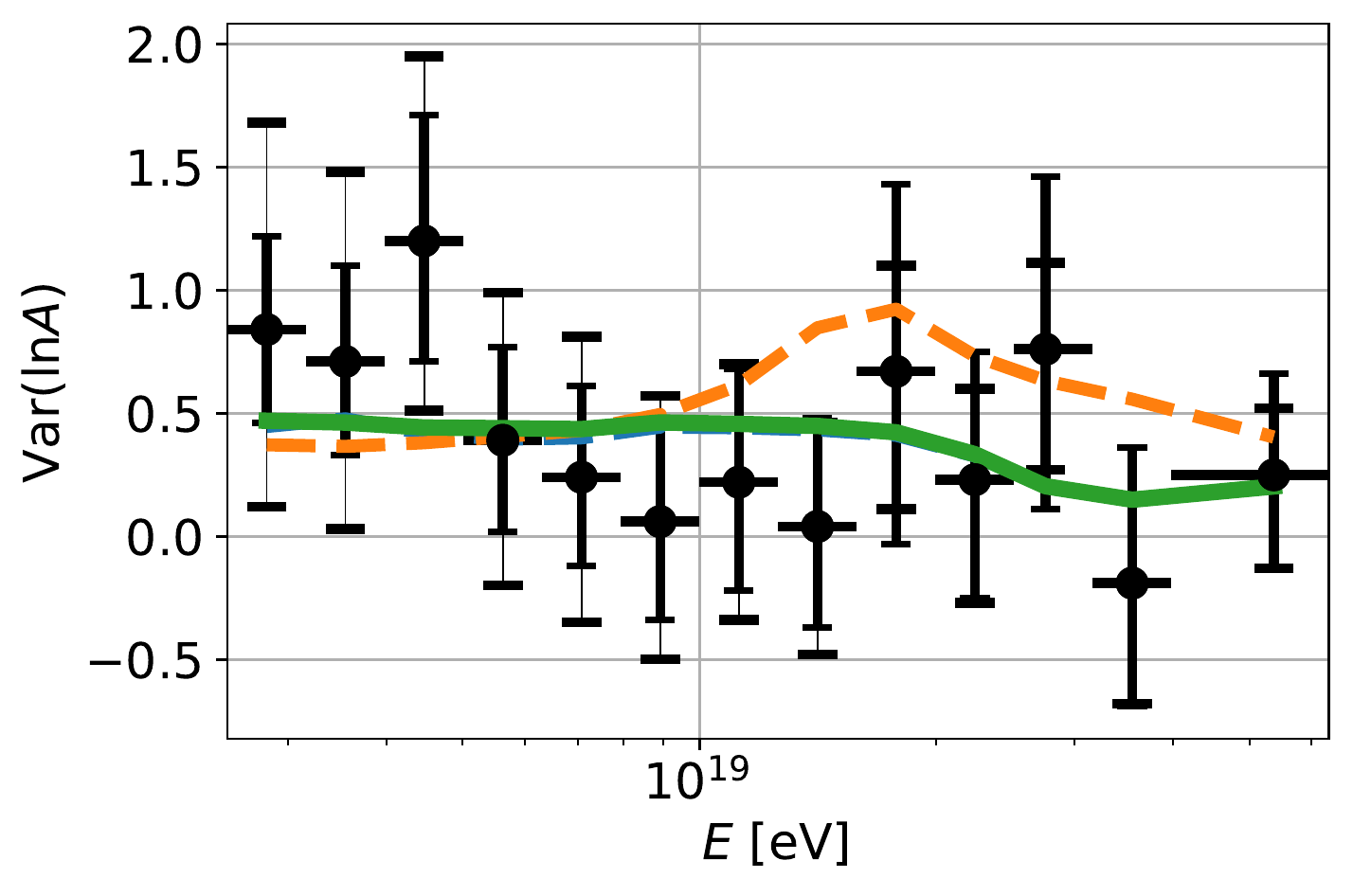}
\includegraphics[width=.45\linewidth]{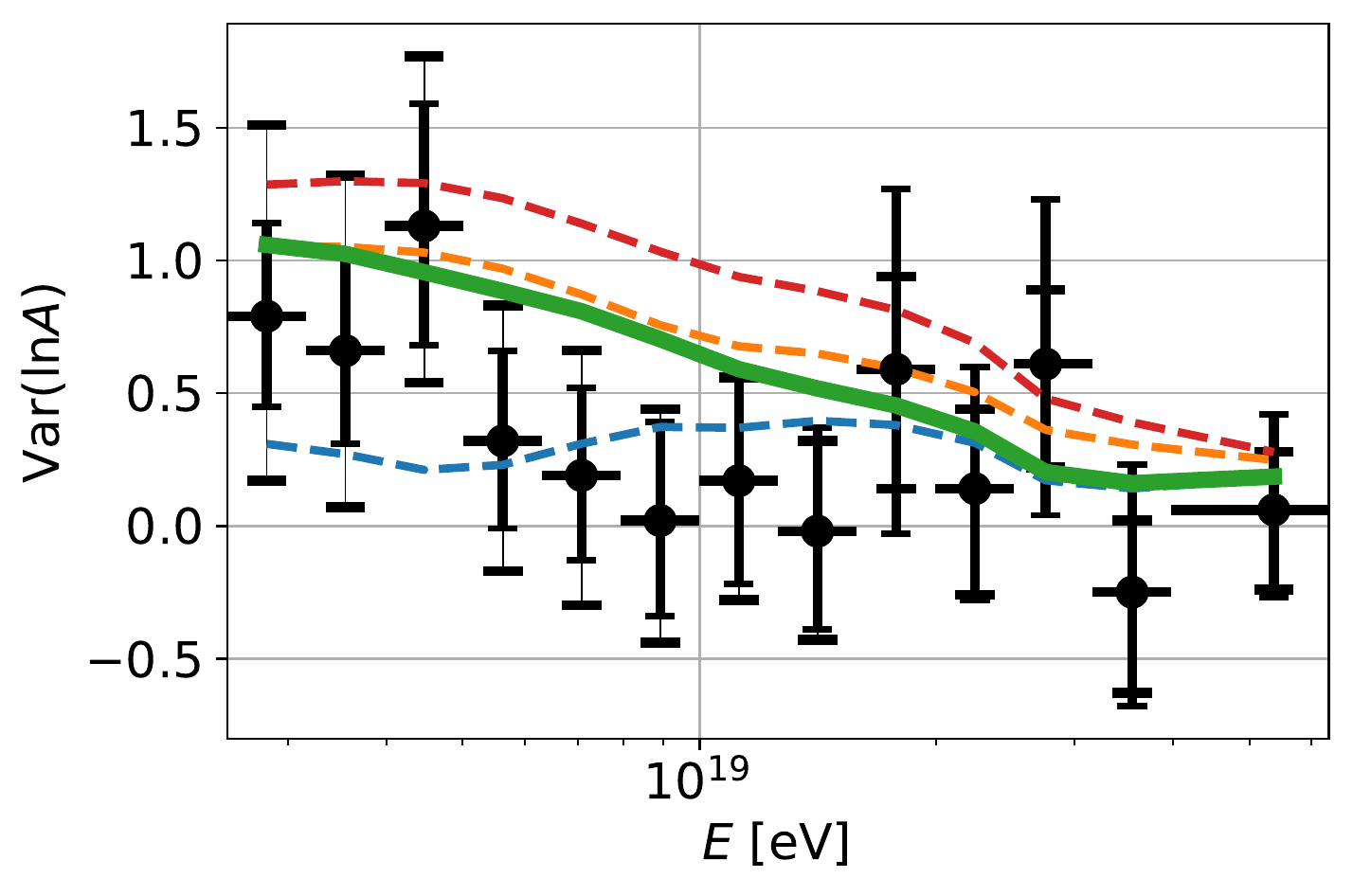}
\caption{Two proof-of-principle fits to the observed energy spectrum \cite{PierreAuger:2020qqz} and the mass composition \cite{Yushkov:2019} (small/large caps of the error bar indicated the statistical/systematical uncertainty) using Fornax A as the single local source. Some of the major differences between the panels are as follows: \\
Scenario A (\emph{left panel}): A different source composition at Fornax A and a 1\,\text{nG} rms EGMF strength. The mass composition data is shown as predicted from the EPOS-LHC model.\\ 
Scenario B (\emph{right panel}): An equal initial composition at all sources and a 0.5\,\text{nG} rms EGMF strength. The mass composition data is shown as predicted from the Sibyll2.3c model.
}
\label{fig:ana_comp}
\end{figure}

In this section we give some illustrative examples of the consequences of the
finite life-time of the sources with respect to the resulting energy spectrum
and the mass composition. Because of the large parameter space of this
model, a fit to all UHECR data is beyond the scope of this work. 
Therefore, we do not use a dedicated fit algorithm to explain these data but
provide different scenarios that show the general consequences and challenges. 
As a result, the proof-of-principle examples shown in Fig.~\ref{fig:ana_comp}
do not explain all data perfectly: We recognize that in particular the
combination of the almost vanishing variance of $\ln A$ at
$7\lesssim E\lesssim 10\,\text{EeV}$ and the rather minor increase of the mean
of $\ln A$ at $6\,\text{EeV}\lesssim E\lesssim 40\,\text{EeV}$ is challenging
to reproduce. However, a dedicated examination of the whole parameter space
would clearly enable an improvement of the fit.
In the following, we only account for a single local source to explain the CRs
above the ankle---even though previous works \cite{Eichmann:2022ias} have shown
that at least a few sources are needed to account for the observed
anisotropy---since this case yields quite illustrative results and keeps the
parameter space rather small. However, we will discuss some consequences of
multiple local sources in Sec.~\ref{sec:disc&concl}. 

We first consider a scenario, hereafter referred to as \emph{Scenario A}, with an individual source modelled similar to Fornax~A, where UHECRs escape diffusively and subsequently propagate through
a strong turbulent EGMF (with a RMS field strength of 1\,nG,
a Kolmogorov spectrum and a coherence length of $0.2\,$Mpc). Moreover, we suppose that all sources eject UHECRs with an initial spectral index of $\alpha=2$. 
Using a finite life-time of $t_{\rm act}=4.1\,\text{Myr}$ where Fornax~A is
currently about in its final stage ($t_{\rm max}\sim -t_{\rm act}$), the left
panel of Fig.~\ref{fig:ana_comp} shows that the observational data can be
explained by the combination of two contributions: the background of
low-luminosity radio galaxies (predominantly below the ankle due to a rather
low acceleration efficiency $g_{\rm acc}=0.15$) and Fornax~A (above the ankle).
However, in this case the initial mass composition of Fornax~A and
the background sources needs to be different: 
While the fraction of hydrogen and helium is about equal, the background (CSF) sources should eject at least about an order of magnitude less of the heavier elements as the individual source (where we used $\sim 10\%$ of carbon, $\sim 1\%$ of neon, $\sim 0.5\%$ of silicon, and $\sim 0.05\%$ of iron). 
Note that, we used the hadronic interaction model \emph{EPOS-LHC} \cite{PhysRevLett.101.171101,PhysRevC.92.034906} to convert the observed $X_{\rm max}$ into $\ln A$, as provided by Ref.~\cite{Yushkov:2019}. But in case of Scenario A a similar fit is in principle possible for any other hadronic interaction model after some adjustment of the initial compositions.
\\
The right panel of Fig.~\ref{fig:ana_comp} shows the results for a different scenario, hereafter referred to as \emph{Scenario B}, where an equal initial composition at all sources is adopted that, however, is in comparison much heavier with $\sim 10\%$ of hydrogen, $\sim 72\%$ of helium, $\sim 14\%$ of carbon, $\sim 3\%$ of neon, $\sim 0.1\%$ of silicon, and $\sim 0.01\%$ of iron. 
Moreover, a weaker rms EGMF strength of $0.5\,\text{nG}$ has been used, so that we had to reduce the source life-time of Fornax A to $t_{\rm act}=1.5\,\text{Myr}$ in order to obtain a proper hardening of the individual CR nuclei spectra at Earth. Another important difference with respect to Scenario A is the use of a somewhat higher acceleration efficiency (with $g_{\rm acc}=0.27$) of the low-luminosity radio galaxies, so that in total their contribution above the ankle becomes significantly higher. Note that with respect to the small level of anisotropy that is observed, a higher (apriori close to isotropic) contribution by the bulk of radio galaxies reduces the need for a strong EGMF or a large number of local sources. 
The composition data in the right panel of Fig.~\ref{fig:ana_comp} have been determined using the interaction model of \emph{Sibyll2.3c} \cite{Riehn2019_sibyll2.3c}, since for  \emph{EPOS-LHC}, that predicts a lighter chemical composition, the fit to the composition data becomes significantly worse for Scenario B.
Apart from some minor discrepancies in the total energy spectrum the two scenarios differ predominantly with respect to their prediction on the chemical composition: Scenario A yields an almost constant $\text{Var}(\ln A)$ throughout the considered energy range, that is in very good agreement with the observations above $5\,\text{EeV}$. But on the downside, the small variance yields a strong increase of $\langle \ln A \rangle$ around 10\,EeV, which leads to some tension with the observational data above 10\,EeV. In contrast, Scenario B shows generally a higher variance of the chemical composition---in particular below about\,10 EeV---yielding some discrepancies with the observational data, but on the upside $\langle \ln A \rangle$ increases less around 10\,EeV, so that it agrees much better to the observational data above 10\,EeV. 
\\
Apart from these differences both scenarios also show several similarities, such as that Fornax~A needs to be rather old, so that its past CR luminosity has been significantly larger. Note that Fornax~A has in comparison to other close-by radio sources already a rather high present jet power. However, it also shows indications of enhanced activity in the past~\cite{Matthews+2018} and in addition, the present jet power estimate can also be enhanced under the assumption of a different radio-to-jet-power correlation, such as the one from Godfrey and Shabala \cite{GodfreyShabala2013}. Moreover, in both fit scenarios a diffusive escape time of about $\tau_{\rm esc}\simeq t_{\rm act}/4$ at 10\,EeV has been adopted. However, the results are hardly sensible to the actual escape process, so that similar fits could also be obtained for an advective scenario with similar values of $\tau_{\rm esc}$. 

\begin{figure}[htbp]
\centering
\includegraphics[width=.55\linewidth]{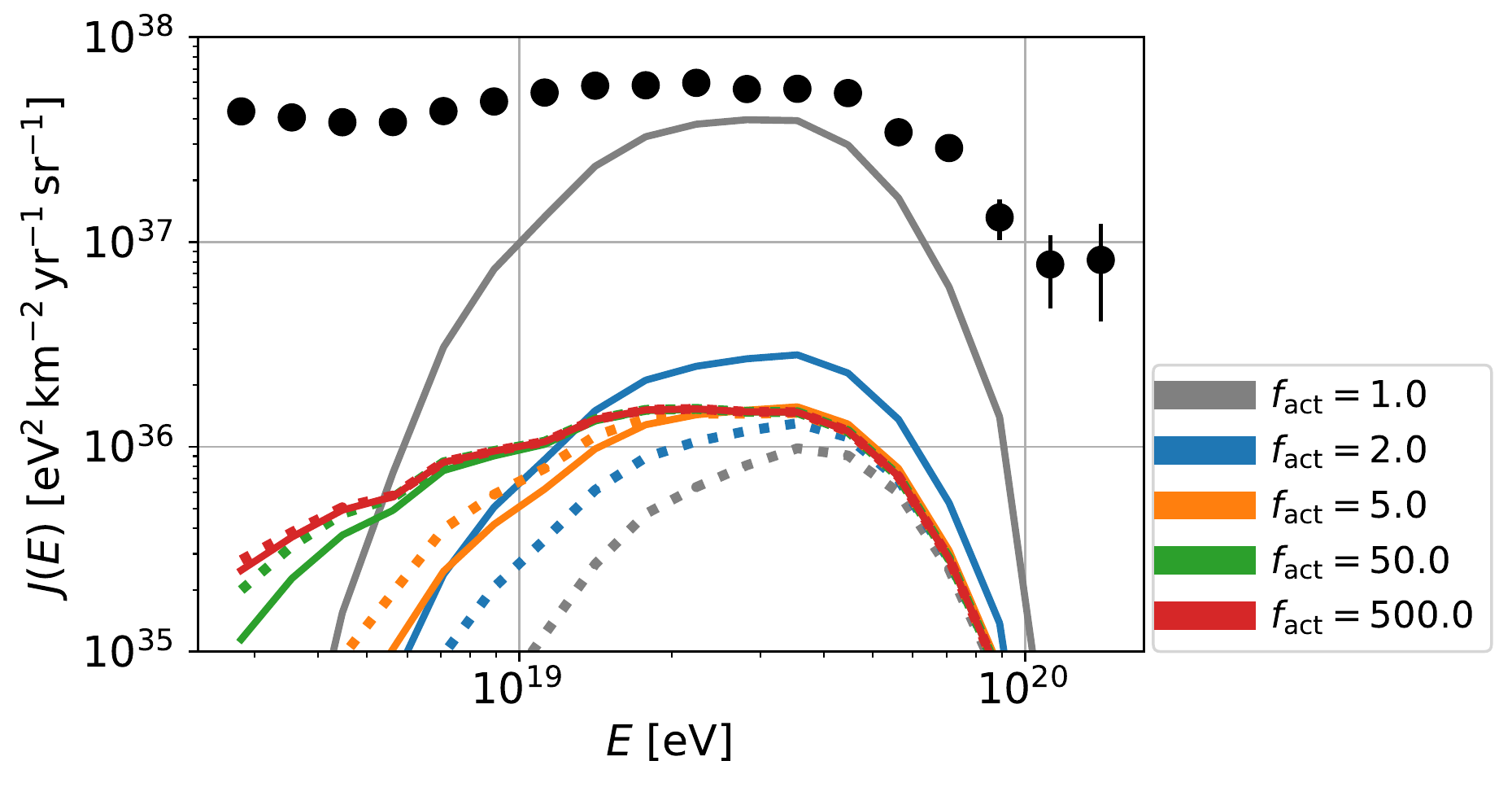}
\includegraphics[width=.42\linewidth]{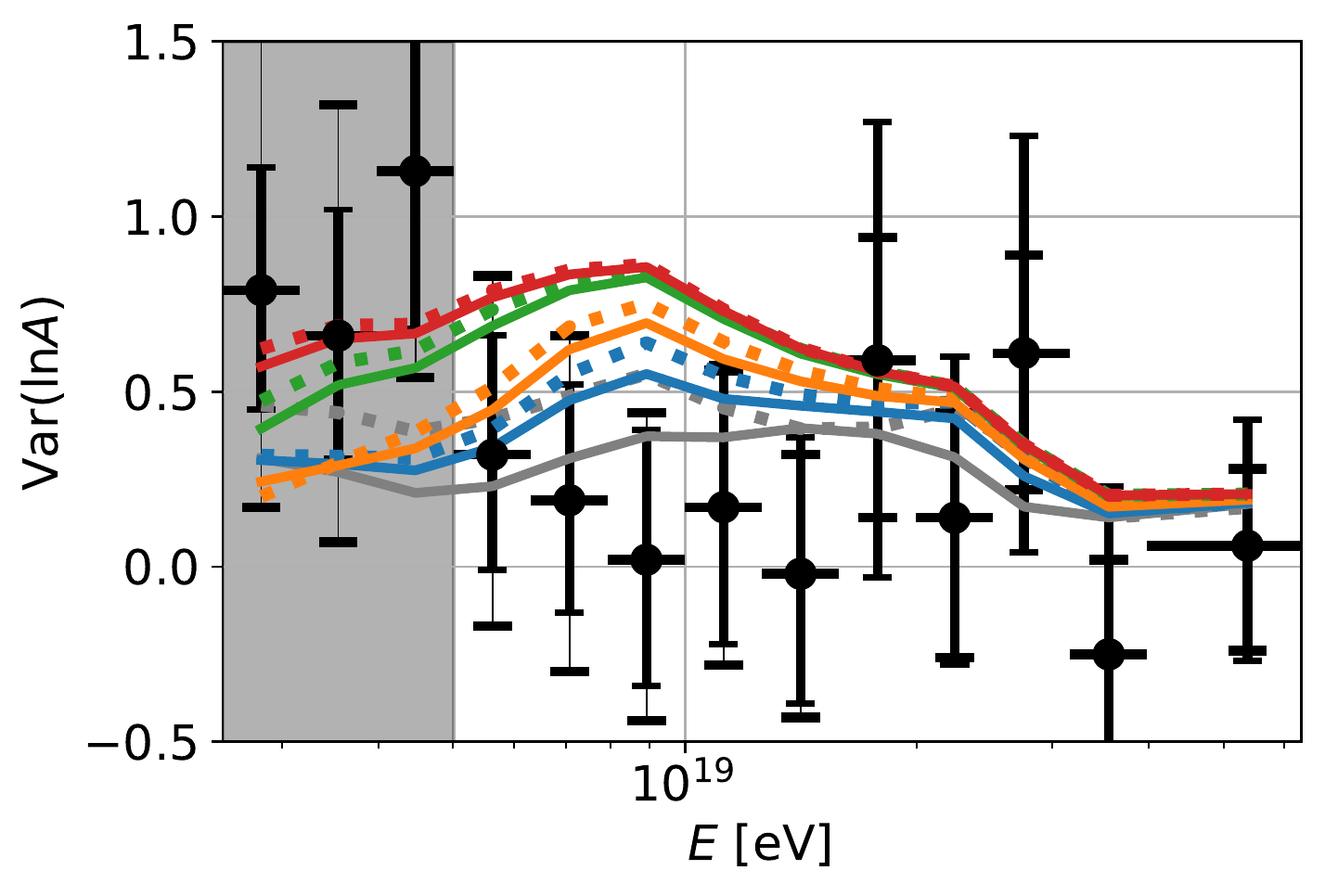}
\caption{The impact of an extended life-time $1.5\,f_{\rm act}\,\text{Myr}$ of Fornax A, where all other parameters are as in Scenario B. In addition to the Gaussian luminosity profile (solid lines) according to Eq.~\ref{eq:escLcr}, the dotted lines indicate the simple case for a step function luminosity profile. \emph{Left:} The energy spectrum of $A=[5,20]$ CR nuclei. \emph{Right:} The variance of $\ln A$ of the Fornax A contribution. Here the grey region indicates where the total $\text{Var}(\ln A)$ will be dominated by the contribution from the bulk of radio galaxies.}
\label{fig:hardeningPlot}
\end{figure}
Generally, it can be seen that taking into account the finite life-time of the
individual local source a quite accurate agreement with the observational data
at energies $\gtrsim 10\,\text{EeV}$ can be obtained: The hard spectra from
the individual initial elements that result from the finite life-time of the
sources---low energy CR are suppressed as they cannot reach the observer in the
given life-time---enable an increase of $\langle \ln A\rangle$ while keeping
$\text{Var}(\ln A)$ small. 
But to obtain the observed CR flux at Earth, even the most powerful local radio sources provide typically not enough CR power in the case of an $1/E^2$ source spectrum, unless theses sources had a significantly higher CR power (of about an order of magnitude) in the past. This is illustrated in the left panel of Fig.~\ref{fig:hardeningPlot} where we modified $t_{\rm act}$ by a factor $f_{\rm act}$, but keep all other parameters of Scenario B fixed---including the time of maximal emission (which is only relevant in case of the Gaussian luminosity profile) of $t_{\rm max}=-1.4\,\text{Myr}$. Hence, using $f_{\rm act}\gg 1$ yields $t_{\rm act}\gg -t_{\rm max}$, so that the past luminosity of the source hardly increases and the resulting flux contribution converges towards the case of a step function luminosity profile, where the source has a constant luminosity during $t_{\rm act}$. In case of $f_{\rm act}\geq 500$ about all CRs manage to reach Earth in time, so that at low energies the spectrum of CR nuclei with $A=[5,\,20]$ show the initial spectral behavior of $1/E^2$. In addition, it can also be seen in Fig.~\ref{fig:hardeningPlot} that nearby sources such as Fornax~A are constrained to have rather
short life-times of only a few Myr, if the EGMF strength is $\lesssim 1\,\text{nG}$, 
otherwise the width of the flux contributions from the individual elements increases leading to an increase of $\text{Var}(\ln A)$ around $10\,\text{EeV}$ by the increasing contribution of heavy nuclei at lower energies.

We verified that also other powerful local sources could in principle be used to explain the data above the ankle. However, this requires some adjustment of the life-time and/or the EGMF strength: 
Supposing a very high EGMF strength of the order of $10\,\text{nG}$, which could be present in the Supergalactic plane, also very nearby sources such as Centaurus A could provide the proper hardening of the individual CR nuclei spectra if their activity time is on the order of a few Myr. More distant sources such as Centaurus B would need a much longer activity time of the order of a few $\times 100\,\text{Myr}$.
Note that on timescales of few Gyr CRs with energies of a few tens of EeV suffer from significant energy losses by the CMB and the EBL, yielding also an upper limit on the possible source distance and relevant activity time, respectively. 
But for significantly smaller timescales energy losses can be neglected in the first place and there is a distinct correlation between the source distance and life-time as well as the EGMF characteristics that enables a proper description of the data: The average propagation delay $\langle t_{\rm del} \rangle$ of CRs with a rigidity of about $5\,\text{EV}$ needs to equal the source life-time, i.e.
\be
t_{\rm act} \sim \langle t_{\rm del} \rangle \simeq 1.2\,\left( \frac{B}{1\,\text{nG}} \right)^2\,\left( \frac{d_{\rm src}}{10\,\text{Mpc}} \right)^2\,\left( \frac{l_{\rm coh}}{1\,\text{Mpc}} \right)\,\text{Myr}\,,
\label{eq:tact-delay}
\ee
assuming the limiting case of small-angle scattering without energy
losses~\cite{Achterberg+1999}. Hence, if the source is much longer active,
$t_{\rm act}\gg \langle t_{\rm del} \rangle$, there is no flux suppression at
low rigidities and in the opposite case,
$t_{\rm act}\ll \langle t_{\rm del} \rangle$, this suppression affects also
the CRs at the highest rigidities leaving not enough CRs to explain the data. 

As a general rule of thumb we can conclude that: The closer the local source,
the shorter needs to be its life-time (or the stronger the EGMF), to provide
a low variance of the mass composition at about 10\,EeV. 

\begin{figure}[htbp]
\centering
\includegraphics[width=.55\linewidth]{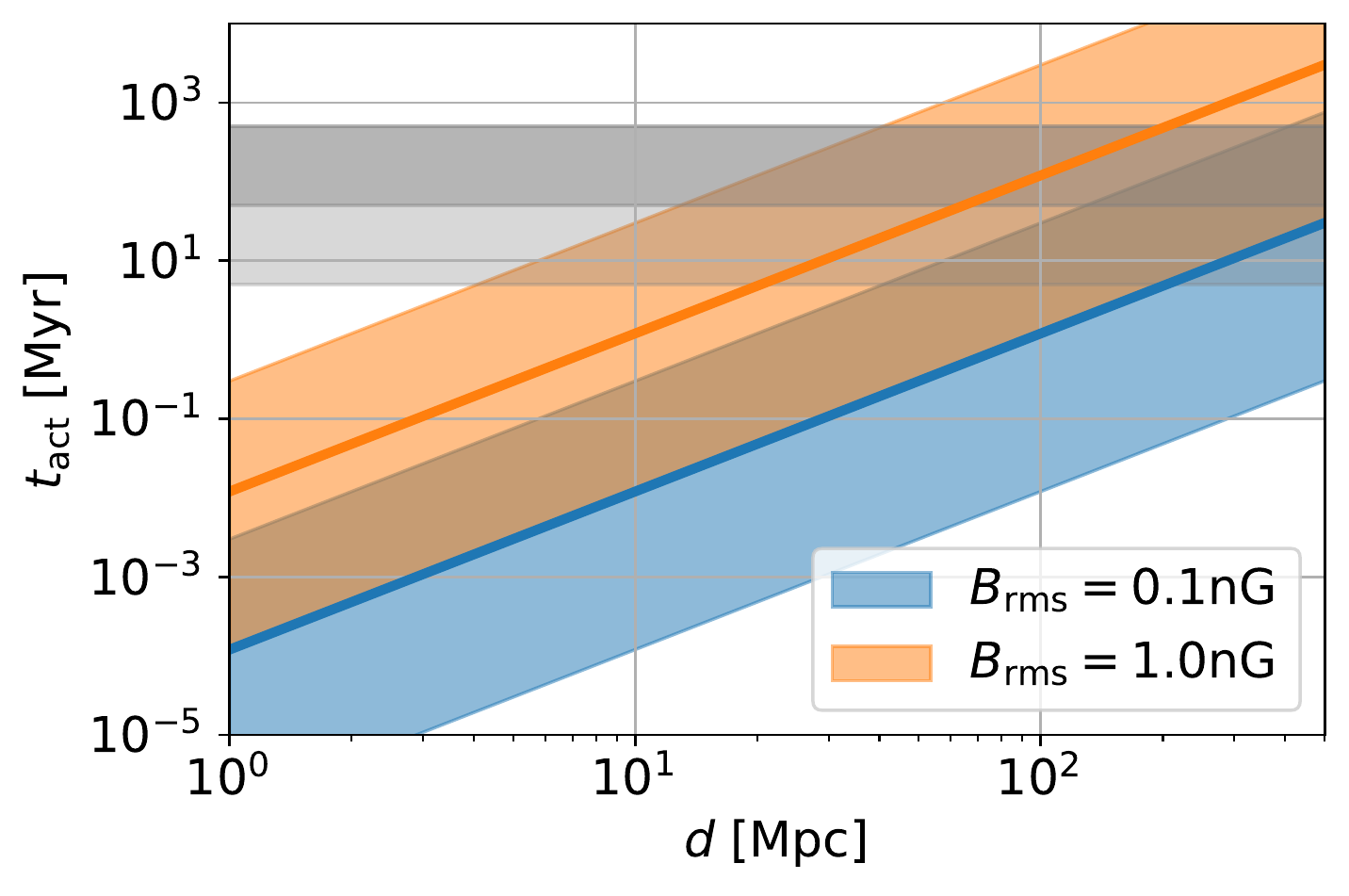}
\caption{The possible activity time of UHECR sources as function of their distance required to obtain a sufficient magnetic suppression at about the ankle according to Eq.~(\ref{eq:tact-delay}). We adopted $l_{\rm coh}=1\,\text{Mpc}$ and two different rms EGMF strengths as indicated by color. Solid line shows the case of $t_{\rm act}=\langle t_{\rm del} \rangle(5\,\text{EV})$ and the colored bands refer to $t_{\rm act}>\langle t_{\rm del} \rangle(50\,\text{EV})$ and $t_{\rm act}<\langle t_{\rm del} \rangle(1\,\text{EV})$. The dark grey band indicates the typical duration of the active phase of radio galaxies.}
\label{fig:tact-dist}
\end{figure}

The Fig.~\ref{fig:tact-dist} shows the consequences of this relation on the possible activity time of UHECR sources dependent on their distance to Earth. Moreover, the grey bands indicate the range of typical outburst and activity timescales of radio galaxies, e.g.~\cite{Jurlin+2020,Konar+2013}. However, also shorter outburst intervals (light grey band) up to $\sim 5\,\text{Myr}$ have been derived from studying X-ray cavities \cite{Vantyghem+2014}.
Local radio sources with distances of a few $\times 10\,\text{Mpc}$ need therefore a rather strong EGMF and close-by sources like Centaurus A, that shows indications of enhanced activity within the last $\sim 100\,\text{Myr}$ \cite{Matthews+2018}, can clearly not provide a sufficiently low variance of the mass composition at about 10\,EeV. Note that in our model we neglect any CR contribution from past activity periods which is in particular at low energies not accurate for $t_{\rm act}\ll 100\,\text{Myr}$, as the quiescent phase is typically smaller than the active phase, i.e. $0.1-10\,\text{Myr}$ \cite{Jurlin+2020,Konar+2013}. However, even for a vanishing duration of the quiescent phase, past activity phases can become negligible if the duration between the different maxima in the luminosity evolution (in case of a Gaussian luminosity profile) becomes sufficient long or the total CR power of the past activity/outburst phases is much smaller than the most recent one. Still, multiple activity or outburst phases can reduce the hardening of the individual CR nuclei spectra and introduce an additional CR contribution at low energies.
%

\section{Discussion and conclusions}
\label{sec:disc&concl}

In this work, we have examined the consequences of a finite life-time of
UHECR sources, using as specific example the case of radio galaxies.
We have used the simulations code CRPropa3~\cite{CRPropa3_2016, CRPropa3.1_2017, CRPropa3.2_2022}
to account for the energy losses and the deflections of UHECRs by the
turbulent EGMF. 
The main motivations to account for the finite life-time of UHECR sources
was to explain the hardness of
the observed UHECR spectra of individual mass groups as a propagation effect
rather than as a characteristics of the ejected source spectra.
This idea has already been investigated in previous works \cite{Harari+2021}. In contrast to this reference, we have
developed for the specific example of radio galaxies a concrete realisation
of this proposal, taking into account both the individual properties of
selected nearby sources as well as the background from the bulk of radio
galaxies at larger distances. On the downside, we neither address the impact
of the finite life-time on the arrival directions, as recently done by
Ref.~\cite{Eichmann:2022ias,Harari+2021}, nor do we perform a proper fit to the
data. Instead we focus on the basic principles, consequences and challenges
of that idea, as summarized in the following:
\begin{enumerate}
    \item[$\bullet$] The finite life-time of the CR sources yields a hardening of their spectra at low rigidities due to the so-called "magnetic horizon" suppression. We showed in Fig.~\ref{fig:ana_comp} that even for a standard acceleration spectrum of $1/E^2$, these hardening effects are in principle sufficient to explain the composition data at the highest energies. 
    \item[$\bullet$] An additional hardening of the CR spectra at low rigidities by $2-m$, where $m$ refers to the spectral index of the turbulence spectrum, can result from the diffusive escape from the source environment. However, this hardening alone is not sufficient for a standard acceleration spectrum to explain the data. 
    \item[$\bullet$] The condition that a present CR source with a given life-time contributes significantly to the observed UHECRs leads to constraints on its size, magnetic field properties and wind speeds, as shown in Fig.~\ref{fig:escTimes}.
    \item[$\bullet$] For a homogeneous distribution of CR sources, there is no hardening of the resulting spectra even for a finite source life-time. Hence, only a few of these (present or past) sources can be responsible for the CRs at the highest energies. 
    \item[$\bullet$] If all source have the same values of the parameters $g_{\rm m}$, $g_{\rm acc}$ and $\alpha$ (which does still allow for different source spectra dependent on the given radio luminosity), an individual CR source at a distance $d_{\rm src}$ needs a CR luminosity $\gtrsim 10^{45}(d_{\rm src}/10\,\text{Mpc})^2\,\text{erg/s}$ at some stage of their evolution, as shown in Fig.~\ref{fig:ind-csf-rel}. 
    Hence, individual sources at distances $\gtrsim 100\,\text{Mpc}$ are not powerful enough to exceed the CR contribution from the bulk of radio sources.
    \item[$\bullet$] A particular combination of the EGMF strength, the source distance, life-time and time of maximal emission is needed to obtain the proper width of the resulting spectra of the individual CR nuclei at Earth. Generally, young sources whose luminosity evolution is currently still increasing do not provide enough UHECRs. Including the previous result of a rather small source distance, we can conclude that the source life-time needs to be $\lesssim 100\,\text{Myr}$. Generally, the closer the local source, the shorter needs to be its life-time (or the stronger the EGMF), as illustrated by Fig.~\ref{fig:tact-dist}. 
    \item[$\bullet$] 
      Although we could explain the observed extreme hard spectra of
      individual mass groups as an effect of the finite life-times of a few dominating UHECR sources, this approach also introduce a few
      challenges:
      (i) the resulting flux contributions of (initial) helium and carbon CRs can become shifted to too low energies to explain the observed spectral behavior at about 10 EeV; (ii) the necessary elongation of the propagation length (due to EGMF deflections) typically yields a sharp cut-off at an energy $\leq 100\,\text{EeV}$; (iii) the observed vanishing variance of $\ln A$ around about 10 EeV is difficult to explain: The need for hard spectra of the individual elements---that sum up to a softer spectrum without any significant dips---leads unavoidably to rather large values of $\diff \langle \ln A \rangle/\diff E$ that challenge the observational data of $\langle \ln A \rangle$ around 10 EeV.
      \item[$\bullet$] If the local EGMF is rather weak ($B_{\rm rms}\ll 1\,\text{nG}$), the typical evolution timescales of radio galaxies indicate that either the recent activity/outburst phase is already too long---especially for nearby sources---or there is the change of a non-vanishing contribution from past activity phases, leading to a too weak  suppression of CRs at energies of about the ankle by the magnetic horizon.
\end{enumerate}
Note that the anisotropy data introduces the need for at least a couple of individual local sources that contribute CRs above the ankle (see e.g. \cite{Eichmann:2022ias}). The necessary number of sources to explain the observed high level of isotropy depends strongly on the actual Galactic and extragalactic magnetic field as well as the source direction and distance. Moreover, it needs to be taken into account that not only the observed dipol anisotropy but in particular the (current) absence of a significant quadrupole anisotropy as well as the intermediate-scale anisotropies above $32\,\text{EeV}$ yield strong constraints on the necessary amount and location of the sources---as e.g.\ shown in Ref.~\cite{Eichmann:2022ias}. But to keep the variance of $\ln A$ almost vanishing at about $10\,\text{EeV}$, all sources need to show a similar spectral behavior of the CR contribution from different types of nuclei, and hence, if the UHECR sources are not standard candles---as indicated by the radio-to-jet power correlation---the amount of relevant sources needs to be small. 
This conclusion is in good agreement with what has been found recently in Ref.~\cite{Ehlert+2022}. Still, the contribution from multiple sources with slightly different spectral properties can become useful to resolve the previously mentioned challenges (i)-(iii), and it relaxes the need for a very high CR luminosity of the individual sources at some stage of their evolution.

\acknowledgments

BE acknowledges support by the DFG grant EI~963/2-1.


\paragraph{Software:} Some of the results in this paper have been derived using the software packages Numpy \cite{Harris+2011}, Scipy \cite{2020SciPy-NMeth}, Pandas \cite{mckinney-proc-scipy-2010}, Matplotlib \cite{Hunter:2007}, Seaborn \cite{Waskom2021} and the CR simulation tool CRPropa\,3~\cite{CRPropa3_2016, CRPropa3.1_2017, CRPropa3.2_2022}.

\appendix

\section{Weighting cosmic rays from sources with a finite life-time}
\label{app}

\subsection{Individual sources}
\label{App:IndSrc}

In the following, we introduce the normalization of UHECR intensity based on the simulation of individual CR candidates from a quasi-instantaneously bursting source at a time $t^\prime$. We use a simulated, initial CR rigidity distribution 
\be
\left( \frac{\diff N}{\diff \Rin}\right)_{\rm sim} = \frac{N_{\rm sim}}{\ln(\hat{R}_{\rm sim}/\check{R}_{\rm sim})\,\Rin}
\ee
in the rigidity range $\Rin\in[\check{R}_{\rm sim},\hat{R}_{\rm sim}]$, where $N_{\rm sim}$ denotes the total number of simulated CRs. Based on sufficient statistics this initial distribution can still be modified after propagation by using the weight 
\be 
w_{\rm R}=\left(\Rin/\check{R}_{\rm sim}\right)^{1-\alpha}\,\exp(\Rin/\hat{R})\,,
\ee
so that we obtain a generic source spectrum with a spectral index $\alpha$ and an exponential cut-off at $\hat{R}<\hat{R}_{\rm sim}$.
The isotropic UHECR intensity at the time $t_{\rm obs}$ at a (spherical) observer with radius $r_{\rm obs}$ is generally given by 
\be
J(E + \Delta E,\,t_{\rm obs})\equiv \frac{\diff N}{\diff A\, \diff E\,\diff t\,\diff\Omega}=\sum_{\epsilon,i}\frac{w_i(\epsilon,\,t_{\rm obs})}{\Delta E\,4\pi\,r_{\rm obs}^2}\,,
\ee
where all CR particles species $i$ and energies $\epsilon\in[E, E + \Delta E]$ are summed using the normalized total weight
\be
w_i(\epsilon,\,t_{\rm obs}) = \frac{1}{4\pi}\,\frac{L_{{\rm in},i}(\epsilon,\,t_{\rm obs})}{E_{{\rm in},i}}\,w_{\rm R}(\epsilon)\,.
\label{eq:totNormW0}
\ee
Here, the initial UHECR luminosity 
\begin{align}
L_{{\rm in},i}(\epsilon,\,t_{\rm obs})&\equiv Z^\prime_i\int_{\check{R}_{\rm sim}}^{\hat{R}_{\rm sim}} \diff \Rin\,\,\Rin \left( \frac{\diff N}{\diff \Rin}\right)_{\rm sim}\,w_{\rm R}(\epsilon)\,\delta(t^\prime-[t_{\rm obs}-t(\epsilon)])\\
&= \frac{\hat{R}_{\rm sim}^{2-\alpha}\,E_{\alpha-1}\left(\hat{R}_{\rm sim}/\hat{R} \right)-\check{R}_{\rm sim}^{2-\alpha}\,E_{\alpha-1}\left(\check{R}_{\rm sim}/\hat{R} \right)}{\hat{R}_{\rm sim}^{2-\alpha}\,E_{\alpha-1}\left(\hat{R}_{\rm sim}/\hat{R} \right)-\check{R}^{2-\alpha}\,E_{\alpha-1}\left(\check{R}/\hat{R} \right)}\,\frac{f_i\,Z^\prime_{i}}{\bar{Z}}\,L_{\rm cr}(t_{\rm obs}-t(\epsilon))
\end{align}
is normalized according to the total CR luminosity $L_{\rm cr}$ in the range $\Rin\in[\check{R},\hat{R}]$ at the time $t_{\rm obs}-t(\epsilon)$, where $t(\epsilon)$ denotes the propagation time of the CR candidate. Further, we have to account for the average initial charge number $\bar{Z}=\sum_i f_i\,\Zin_i$ of the mass composition with a normalized abundance $f_i$ of the initial charge number $\Zin_i$. Here, $E_{n}(x)$ denotes the exponential integral function. In addition, the total initial UHECR energy is determined by
\begin{align}
E_{{\rm in},i} &\equiv Z^\prime_i\int_{\check{R}_{\rm sim}}^{\hat{R}_{\rm sim}} \diff \Rin\,\,\Rin \left( \frac{\diff N}{\diff \Rin}\right)_{\rm sim}\,w_{\rm R}(\epsilon) \\
&= \frac{\Zin_i\,N_{\rm sim}\,\left[\hat{R}_{\rm sim}^{2-\alpha}\,E_{\alpha-1}\left(\hat{R}_{\rm sim}/\hat{R} \right)-\check{R}_{\rm sim}^{2-\alpha}\,E_{\alpha-1}\left(\check{R}_{\rm sim}/\hat{R} \right)\right]}{\ln(\hat{R}_{\rm sim}/\check{R}_{\rm sim})\,\check{R}_{\rm sim}^{1-\alpha}}\,
\end{align}
so that the normalized total weight (\ref{eq:totNormW0}) yields
\be
w_i(\epsilon,\,t_{\rm obs})= \frac{f_i\,L_{\rm cr}(t_{\rm obs}-t(\epsilon))\,\ln(\hat{R}_{\rm sim}/\check{R}_{\rm sim})\,\Rin{}^{1-\alpha}\,\exp\left(-\Rin/\hat{R}\right)}{4\pi\,\bar{Z}\,N_{\rm sim}\,\left[\hat{R}_{\rm sim}^{2-\alpha}\,E_{\alpha-1}\left(\hat{R}_{\rm sim}/\hat{R} \right)-\check{R}_{\rm sim}^{2-\alpha}\,E_{\alpha-1}\left(\check{R}_{\rm sim}/\hat{R} \right)\right]}\,.
\label{eq:weight_ind}
\ee

\subsection{Continuous sources}
\label{App:CSF}

Based on the simulation of individual CR candidates from a uniform 1D distribution of (discrete) sources in light travel distance between $\check{d}_{\rm L}(t_{\rm obs})$ and $\hat{d}_{\rm L}$, the isotropic UHECR intensity is generally given by 
\begin{align}
J_{\rm csf}(E + \Delta E,\,t_{\rm obs})&=\frac{c}{4\pi} \sum_{\epsilon,i,j}\Delta z_j\,\,\left| \frac{\diff t}{\diff z}\, \right|\,\frac{\Rin\,\Psi_{i}(\Rin,\,z_j)}{\Delta E} \\
&\simeq \frac{1}{4\pi} \sum_{\epsilon,i,j}\frac{\bigl(\hat{d}_{\rm L}-\check{d}_{\rm L}(t_{\rm obs})\bigr)}{N_{\rm sim}}\,\frac{\Rin\,\Psi_{i}(\Rin,\,z_j)}{\Delta E}\,.
\label{eq:Jcsf_app}
\end{align}
Note, that we need a sufficient number of statistic in the last step of calculation in order to approximate that 
\be
\Delta z_j = \frac{\Delta d_{{\rm L},j}}{H_0^{-1}c}\,\mathcal{E}(z_j)(1+z_j) \simeq \frac{\bigl(\hat{d}_{\rm L}-\check{d}_{\rm L}(t_{\rm obs})\bigr)}{N_{\rm sim}\,H_0^{-1}c}\,\mathcal{E}(z_j)(1+z_j)\,,
\ee
where we introduced the dimensionless Hubble parameter $\mathcal{E}(z_j) = H(z_j)/H_0$.

\bibliographystyle{JHEP} 
\addcontentsline{toc}{section}{References}
\bibliography{references}

\end{document}